\documentclass[12pt]{article}
\usepackage{hyperref}
\usepackage[top=2cm,bottom=3cm,left=2cm,right=2cm]{geometry}
\usepackage{graphicx}
\usepackage{amsmath}
\usepackage{amssymb} 
\usepackage{bm} 

\numberwithin{equation}{section}
\numberwithin{figure}{section}

\def\eq#1{(\ref{eq:#1})}

\newcommand{\Tr}{\mathop{\rm Tr}\nolimits}
\def\d{\partial}
\def\eps{\epsilon}

\newcommand{\inverttriangle}{%
               \mathrel{\raisebox{.1em}{%
               \reflectbox{\rotatebox[origin=c]{180}{$\triangle$}}}}\!}

\def\H{\mathcal{H}}
\def\Phat{\widehat{\mathcal{P}}}
\def\P{\mathcal{P}}
\def\Ppre{\mathcal{P}_\mathrm{pre}}
\def\deltahat{\widehat{\delta}}
\def\L{\mathcal{L}}

\def\M{{\bf M}}
\def\llambda{{\bm \lambda}}
\def\xxi{{\bm \xi}}
\def\f{{\bf f}}

\begin{document}

\begin{titlepage}
	
	\hfill \today
	\begin{center}
		\vskip 2cm
		
		{\Large \bf {Conserved charges and $L_\infty$ algebras}
		}
		
		\vskip 0.5cm
		
		\vskip 1.0cm
		{\large {Vin\'{\i}cius Bernardes$^{1}$, Theodore Erler$^{1}$, and Atakan Hilmi F{\i}rat$^{2}$ }}
		
		\vskip 0.5cm
		
		{\em  \hskip -.1truecm
			$^{1}$
			CEICO, FZU - Institute of Physics of the Czech Academy of Sciences \\
			No Slovance 2, 182 21, Prague 8, Czech Republic
			\\
			\vskip 0.5cm
			$^{2}$
			Center for Quantum Mathematics and Physics (QMAP),
			Department of Physics \& Astronomy, \\
			University of California, Davis, CA 95616, USA
			\\
			\vskip 0.5cm
			\tt \href{mailto:viniciusbernsilva@gmail.com}{viniciusbernsilva@gmail.com},
		\href{mailto:tchovi@gmail.com}{tchovi@gmail.com},
		 \href{mailto:ahfirat@ucdavis.edu}{ahfirat@ucdavis.edu} \vskip 5pt }
		
		\vskip 2.0cm
		{\bf Abstract}
		
	\end{center}
	\vskip 0.25cm
	\noindent
	\begin{narrower}
		\baselineskip15pt
		We give a formula for conserved charges in an arbitrary Lagrangian field theory expressed in the framework of $L_\infty$ algebras. The formula is determined by the theory's $L_\infty$ data alone, without reference to the derivative structure of the Lagrangian. Therefore conserved charges can be computed in nonlocal models, such as string field theory, where conventional methods break down. The formula also gives the correct expression for the Hamiltonian of general relativity as a surface integral of the Brown-York stress tensor. Related computations in Yang-Mills theory suggest that spatial boundaries are dealt with in a natural fashion.  
	\end{narrower}
\end{titlepage}

\tableofcontents
\baselineskip15pt

\section{Introduction}

In recent work \cite{Bernardes} we argued that any field theory formulated in terms of a cyclic $L_\infty$ algebra has a formula for the symplectic structure on its phase space:
\begin{equation}\Omega = \frac{1}{2}\omega\big(\delta\Phi,[Q_\Phi,\sigma]\delta\Phi\big),\label{eq:Omega}\end{equation}
where $Q_\Phi$ is the kinetic operator around a solution $\Phi$, $\delta$ is the exterior derivative on the space of solutions, $\omega$ is the Batalin-Vilkovisky (BV) inner product, and $\sigma$ is an operator called the {\it sigmoid}. In this paper we use this formula to construct conserved charges associated to symmetries of the action. In a companion paper \cite{Bernardes5} we construct the Poisson bracket, providing all the basic ingredients for Hamiltonian mechanics. The major advantage of this approach is that it does not rely on the derivative structure of the Lagrangian. Therefore it applies to nonlocal field theories where the canonical construction of coordinates and momenta may not be possible. The ultimate goal is to compute phase space observables in string theory, such as energy or black hole entropy~\cite{Wald2}, in a framework that is fully nonperturbative in $\alpha'$.

Our discussion is based on the $L_\infty$ algebra approach to field theory \cite{Hohm,Jurco}. The algebraic structure guides the construction and provides a universal language for expressing the formulas and demonstrating their consistency. It is also natural in the context of string field theory (see \cite{deLacroix,Erler,Erler2,Erbin,Maccaferri,Sen} for introduction), which up to now has been the focus of applications. A~recent series of papers used the symplectic structure to compute the energy of rolling tachyon solutions~\cite{Bernardes2}, the momentum of sliding D0-branes \cite{Bernardes3}, and the energy of an electrified D-brane~\cite{Bernardes4}. In upcoming work \cite{Bernardes6} we compute the mass of a D0-brane from the central charge of the spontaneously broken Poincar{\'e} algebra in 26 dimensions. In the present paper we do not discuss string field theory. We will illustrate the formalism in simpler and more conventional field theory models. 

In the language of $L_\infty$ algebras, a symmetry transformation can be expressed in the form 
\begin{equation}
\mathcal{L}_\xi\Phi = \sum_{n= 0}^\infty \frac{1}{n!}\Xi_n(\underbrace{\Phi,\cdots,\Phi}_{n\text{ times}}),\label{eq:xiintro}
\end{equation}
where $\Xi_n$ are a sequence of higher order products with the appropriate properties. The conserved charge can be written 
\begin{align}
F_\xi = \sum_{n=0}^\infty \sum_{k=0}^n \frac{1}{(n+1)!}\left(\begin{matrix} n \\ k\end{matrix}\right) \omega\bigg(\Phi,\ & \Xi_{k+1}\Big(\sigma L_{n-k}(\underbrace{\Phi,\cdots,\Phi}_{n-k\ \mathrm{times}}),\underbrace{\Phi,\cdots,\Phi}_{k\ \mathrm{times}}\Big)\nonumber\\
&\ \ \  -L_{k+1}\Big(\sigma \Xi_{n-k}(\underbrace{\Phi,\cdots,\Phi}_{n-k\ \mathrm{times}}),\underbrace{\Phi,\cdots,\Phi}_{k\ \mathrm{times}}\Big)\bigg),\label{eq:Fvintro}
\end{align}
where $\sigma$ is the sigmoid and $L_n$ are the higher order products which define the classical action. To appreciate the structure, it is worth noticing that if the sigmoid were replaced by $1$,  this expression would vanish because the products of the symmetry commute, in the appropriate sense, through the products of the action. The conserved charge is created as the sigmoid prevents the products from reordering and canceling each other. Later we describe another form of the conserved charge as a total derivative on spacetime which is often more convenient for computation. 

This paper is organized as follows. Section \ref{sec:formalism1} gives the formal derivation of conserved charges. To prepare for this we describe the action and its $L_\infty$ gauge invariance; we introduce the notion of pre-phase space, phase space, the sigmoid, and the symplectic structure; and finally, we describe symmetries of the action and gauge invariant observables. Conserved charges are derived by expressing the contraction of a symmetry with the symplectic structure as an exact 1-form on phase space. We give an expression for the conserved charge which is manifestly localized in time and, if expanded explicitly in terms of the products of the symmetry and the action, reproduces \eq{Fvintro}. A second form of the conserved charge is given as a total derivative on spacetime with a specific regularization of contributions from temporal boundaries. In section \ref{sec:coalgebra} we present these results using the coalgebra description of $L_\infty$ algebras. This gives perspective on the origin of the formalism in section \ref{sec:formalism1} and some additional tools for calculation.  In section \ref{sec:examples} we compute the conserved charge in several examples. The first and simplest is the stress-energy tensor of Klein-Gordon theory in flat space. To test the formalism in the presence of nonlocality, we derive the Hamiltonian of $p$-adic string theory \cite{Brekke} and show that it agrees with the energy functional derived as a derivative expansion by Moeller and Zwiebach~\cite{Moeller2}. The next examples are concerned with the derivation of charges in the presence of spatial boundaries. Though we do not give a systematic treatment of spatial boundaries in this paper, we find that the formula nevertheless appears to account for the needed boundary contributions.  We illustrate this with the derivation of surface charges in Yang-Mills theory and general relativity, as well as Killing charges of Yang-Mills theory in the presence of a spatial boundary. We end with some concluding remarks.

\subsubsection*{Conventions}

We assume that the products of the $L_\infty$ algebra have odd parity, are graded commutative, and carry grade $+1$. All commutators are graded according to whether the object is commuting or anticommuting. The metric is assumed to have mostly positive signature.

\section{Formalism}
\label{sec:formalism1}

In this section we derive the conserved charges associated to symmetries of the action in the language of $L_\infty$ algebras. This requires a bit of formal preparation, concerning the formulation of the action and its gauge invariance, the symplectic structure, symmetries of the action, and gauge invariant observables. We have tried to simplify the presentation to avoid heavy reliance on the mathematics of $L_\infty$ algebras. For those comfortable with these objects, however, the coalgebra presentation is quite useful, and is summarized in section~\ref{sec:coalgebra}. A derivation of the Hamiltonian by a different approach appears in \cite{Ali}. Related discussion of Noether's theorem in the context of homotopy algebras appears in \cite{Konosu}. 

Our results in principle apply to nonlocal field theories. However, we assume that the correlations in the system are at least somewhat localized. A counterexample is the scalar field theory action 
\begin{equation}
S = -\frac{1}{2}\int d^4 x\, \d_\mu\phi \d^\mu\phi -\frac{1}{4}\left(\int d^4 x\, \phi^2\right)^2 \, .
\end{equation}
The quartic vertex couples the scalar field without any suppression over infinite distances and intervals of time. We will not try to be precise about which kind of nonlocal kernels are allowed. Calculations of the symplectic structure \cite{Bernardes,Bernardes2,Bernardes3,Bernardes4} show that the formalism can be applied to nonlocal models such as $p$-adic string theory, Wilsonian effective field theories, and string field theory. 

\subsection{$L_\infty$ action}

We consider a classical action of the form
\begin{equation}
S = -\frac{1}{2}\omega(\Phi,Q\Phi)-\sum_{n=2}^\infty \frac{1}{(n+1)!}\omega\big(\Phi,L_n(\underbrace{\Phi,\cdots,\Phi}_{n\ \mathrm{times}})\big).\label{eq:S}
\end{equation}
The ingredients are:
\begin{itemize}
\item $\Phi$ is the dynamical field. It is an element of a vector space $\H$ which has two gradings: an integer cohomological grading and an even/odd parity describing whether elements are commuting or anticommuting. The dynamical field is grade zero and commuting.
\item $\omega$ is the {\it BV inner product}. It is a graded antisymmetric, nondegenerate bilinear form on $\H$. It  carries grade $-1$ and anticommutes.
\item The $L_n$s for $n\geq 1$ are the {\it $L_\infty$ products}. They are graded symmetric multilinear maps from $n$ copies of $\H$ into $\H$. They carry grade 1 and anticommute. In addition, they satisfy a hierarchy of quadratic identities called {\it $L_\infty$ relations}, the first of which says that the operator $L_1=Q$ is nilpotent: $Q^2=0$. Finally, they are conserved through the BV inner product. This property is often referred to as {\it cyclicity}. 
\end{itemize}
The objects $(\H,\omega,L_n)$ together define a {\it cyclic $L_\infty$ algebra}. More detail on cyclic $L_\infty$ algebras  following our notation can be found in \cite{Erler}. Original sources are~\cite{Zwiebach,Lada,Lada2}. An important consideration is whether the products are cyclic. Cyclicity can be viewed as a generalization of integration by parts, and typically holds only if boundary contributions can be argued to vanish. We will assume that the products and spatial boundary conditions have been defined so as to ensure that the products are cyclic except  perhaps for contributions from temporal boundaries. This is related to the problem of providing the correct boundary terms for the action, such as the Gibbons-Hawking-York (GHY) term in general relativity \cite{York,Gibbons,Harlow,Bentabol}. For now we assume that this problem has been solved and do not discuss it in further detail. However, we argue in subsection \ref{subsec:surface} that the solution may not be needed to derive the conserved charge in the correct form in the presence of spatial boundaries. 

The space of classical fields $\Phi$ is a linear subspace of $\H$ which we denote as $\Phat$. We consider a complex of differential forms on $\Phat$ whose exterior derivative is denoted $\deltahat$. We assume that $\deltahat$ anticommutes and carries grade zero. The Lie derivative along a vector field $\xi$ in $\Phat$ will be denoted~$\L_\xi$. Computing the exterior derivative of the action gives
\begin{equation}\deltahat S = \omega(\deltahat\Phi,q_\Phi),\label{eq:deltaS}\end{equation}
where $q_\Phi$ is the {\it Euler-Lagrange state}
\begin{equation}
q_\Phi = Q\Phi + \sum_{n=2}^\infty\frac{1}{n!}L_n(\underbrace{\Phi,\cdots,\Phi}_{n\ \mathrm{times}}).\label{eq:E}
\end{equation}
This is an anticommuting element of $\H$ at grade 1. When computing the exterior derivative we followed common practice of ignoring temporal boundary terms generated through the use of cyclicity, as these are typically not relevant to the variational principle. The classical equations of motion are 
\begin{equation}
q_\Phi = 0.
\end{equation}
The exterior derivative of the Euler-Lagrange state gives 
\begin{equation}
\deltahat q_\Phi = - Q_\Phi \deltahat\Phi,\label{eq:deltaE}
\end{equation}
where $Q_\Phi$ is the {\it kinetic operator} (around $\Phi$)
\begin{equation}
Q_\Phi A = QA + \sum_{n=1}^\infty \frac{1}{n!}L_{n+1}(\underbrace{\Phi,\cdots,\Phi}_{n\ \mathrm{times}},A),
\end{equation}
which is an anticommuting operator on $\H$ at grade 1. It determines the spectrum of linearized fluctuations around~$\Phi$. The kinetic operator is cyclic up to temporal boundary terms,
\begin{equation}\omega(Q_\Phi A,B) + (-1)^{|A|}\omega(A,Q_\Phi B)=0,\end{equation}
due to cyclicity of $L_n$s. We write $|A|$ which is even or odd if $A$ is commuting or anticommuting. Further we have the Noether identity,
\begin{equation}
Q_\Phi q_\Phi = 0,\label{eq:QE}
\end{equation}
which follows from $L_\infty$ relations. Computing the exterior derivative of $Q_\Phi$ further introduces a 2-product around $\Phi$ and so forth for higher products. These satisfy $L_\infty$ relations, but we will not need to discuss them. 

There is a useful way to rewrite the action in terms of the Euler-Lagrange state \cite{Erler3}. This trick avoids the need to expand the action term-by-term using the $L_\infty$ products and is also useful for expressing the conserved charges. The idea is to introduce a field $\Phi(s)$, where $s\in [0,1]$ is a parameter, and impose boundary conditions
\begin{equation}\Phi(0) = 0,\ \ \ \Phi(1) = \Phi,\end{equation}
where $\Phi$ is the dynamical field. The action can be written as
\begin{equation}
S = -\int_0^1 ds\,\omega\left(\frac{\d\Phi(s)}{\d s},q_{\Phi(s)}\right).
\end{equation}
Though it is not manifest, the integrand is a total derivative in $s$, so the action only depends on the value of $\Phi(s)$ at $s=1$. This can be seen by computing the exterior derivative:
\begin{align}
\deltahat S & = \int_0^1 ds\left[\omega\left(\frac{\d \deltahat\Phi(s)}{\d s}, q_{\Phi(s)}\right)-\omega\left(\frac{\d\Phi(s)}{\d s},Q_{\Phi(s)}\deltahat \Phi(s)\right)\right]\nonumber\\
& = \int_0^1 ds\left[\omega\left(\frac{\d \deltahat\Phi(s)}{\d s}, q_{\Phi(s)}\right)+\omega\left(Q_{\Phi(s)}\frac{\d\Phi(s)}{\d s},\deltahat \Phi(s)\right)\right]\nonumber\\
& = \int_0^1 ds\left[\omega\left(\frac{\d \deltahat\Phi(s)}{\d s}, q_{\Phi(s)}\right)+\omega\left(\deltahat \Phi(s),Q_{\Phi(s)}\frac{\d\Phi(s)}{\d s}\right)\right]\nonumber\\
& = \int_0^1 ds \frac{\d}{\d s}\omega\big(\deltahat\Phi(s), q_{\Phi(s)}\big)\nonumber\\
& = \omega\big(\deltahat\Phi, q_{\Phi}\big).\label{eq:deltaS2}
\end{align}
In the first line we used \eq{deltaE}, in the second line we used cyclicity of $Q_\Phi$ (ignoring temporal boundary terms), in the third we used graded antisymmetry of $\omega$, and in the fourth line we pulled out a total derivative with respect to $s$. Integrating the total derivative reproduces the expected variation of the action, \eq{deltaS}. In particular, we see that changes of $\Phi(s)$ which do not change $\Phi$ leave the action invariant. Therefore the action only depends on $\Phi$. This trick for writing the action in terms of the equations of motion appears also in \cite{Rossi} and appendix A.3 of~\cite{Barnich}. The idea can be traced back to the proof of the Poincar{\'e} lemma.\footnote{We thank Igor Khavkine for pointing this out.}

The $L_\infty$ action has gauge symmetry. This means that $\Phat$ is foliated by submanifolds representing gauge equivalent field configurations. These submanifolds are generated by vector fields which implement the gauge transformations. The Lie derivative along such a vector field (denoted~$\Lambda$) takes the form
\begin{equation}
\L_\Lambda \Phi = Q_{\Phi}\lambda_\Phi + \Lambda_\Phi q_\Phi, \label{eq:gauge}
\end{equation}
where $\lambda_\Phi\in\H$ is called the {\it gauge parameter} and $\Lambda_\Phi:\H\to\H$ is called the {\it gauge operator}. Both are anticommuting and carry grade $-1$. The gauge operator should be defined consistently with spatial boundary conditions to ensure cyclicity,
\begin{equation}\omega(\Lambda_\Phi A,B) + (-1)^{|A|}\omega(A,\Lambda_\Phi B)=0,\end{equation}
up to temporal boundary terms. We may describe $\lambda_\Phi$ and $\Lambda_\Phi$ in terms of products $\Lambda_n$,
\begin{subequations}
\begin{align}
\lambda_\Phi & = \sum_{n=0}^\infty \frac{1}{n!}\Lambda_n(\underbrace{\Phi,\cdots,\Phi}_{n\text{ times}}),\\
\Lambda_\Phi A & = \sum_{n=0}^\infty \frac{1}{n!}\Lambda_{n+1}(\underbrace{\Phi,\cdots,\Phi}_{n\text{ times}},A).
\end{align}
\end{subequations}
The gauge parameter generates an ``ordinary'' gauge transformation, while the gauge operator generates a so-called ``trivial'' gauge transformation that vanishes on-shell. We consider these contributions at the same time as part of the data which define the vector field $\Lambda$. The gauge parameter and gauge operator can in general depend on the field $\Phi$. In this way, the gauge transformation generates a flow which is naturally defined locally at each point in $\Phat$. Gauge transformations form a closed algebra as determined by the Lie bracket of vector fields on $\Phat$. Since we allow trivial gauge transformations and field-dependent gauge parameters, the gauge algebra generalizes the one discussed in \cite{Mamade}. Let us demonstrate that the action is gauge invariant. Using \eq{deltaS} we have
\begin{equation}
\L_\Lambda S = -\omega(\L_\Lambda\Phi,q_\Phi) = -\omega(Q_\Phi\lambda_\Phi,q_\Phi)-\omega(\Lambda_\Phi q_\Phi,q_\Phi).\label{eq:LLambdaS}
\end{equation}
The first term vanishes after using cyclicity and recalling that $Q_\Phi$ annihilates the Euler-Lagrange state. In the second term we note
\begin{equation}
\omega(\Lambda_\Phi q_\Phi,q_\Phi)=\omega( q_\Phi,\Lambda_\Phi q_\Phi) = -\omega(\Lambda_\Phi q_\Phi,q_\Phi),
\end{equation}
where in the first equality we used cyclicity of $\Lambda_\Phi$ and the second equality graded antisymmetry of $\omega$. Therefore the second term must be zero, implying
\begin{equation}
\L_\Lambda S = 0.
\end{equation}
The action is gauge invariant as expected.

\subsection{Covariant phase space}

The space of fields $\Phi$ satisfying the equations of motion
\begin{equation}q_\Phi = 0,\ \ \ \Phi\in \Ppre\subseteq \Phat\end{equation}
is called the {\it pre-phase space}, denoted $\Ppre$. Pre-phase space is a nonlinear submanifold of $\Phat$. We consider a complex of differential forms on pre-phase space whose exterior derivative is denoted by $\delta$. We assume that $\delta$ is anticommuting and carries grade zero. Taking the exterior derivative of the Euler-Lagrange state implies that the differential $\delta\Phi$ is annihilated by the kinetic operator,
\begin{equation}Q_\Phi \delta\Phi = 0.\label{eq:QdeltaPhi}\end{equation}
This means that $\delta\Phi$ represents the set of solutions to the linearized equations of motion around the background~$\Phi$. The spectrum of fluctuations of a vacuum depends on the vacuum, which means $\delta\Phi$ will depend on~$\Phi$. 

\begin{figure}[t]
	\centering
	\includegraphics[scale=1.2]{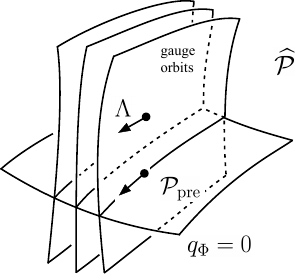}
	\caption{\label{fig:ConsL1} The pre-phase space $\Ppre$ is a subspace of the space of fields $\widehat{\P}$ which satisfies the equations of motion $q_\Phi=0$. Both $\widehat{\P}$ and $\Ppre$ are foliated by gauge orbits generated by vector fields $\Lambda$. Solutions equated through the flow of these vector fields define the phase space $\P$.} 
\end{figure}

Gauge transformations map solutions into solutions. This means that vector fields which generate gauge transformations on $\Phat$ descend to tangent vectors on $\Ppre$. The Lie derivative of a solution $\Phi$ along such a tangent vector (again denoted $\Lambda$) takes the form
\begin{equation}\L_\Lambda\Phi = Q_\Phi \lambda_\Phi,\ \ \ \Phi\in \Ppre.\label{eq:gaugePpre}\end{equation}
The trivial term from \eq{gauge} drops out because the Euler-Lagrange state vanishes on-shell. Gauge transformations define a physical equivalence between solutions. The set of physically inequivalent classical solutions defines {\it phase space}, denoted $\P$. See figure \ref{fig:ConsL1}. This concept of phase space is the foundation for the covariant phase space formalism \cite{Crnkovic,Crnkovic2,Zuckerman,Lee}. For reviews see~\cite{Khavkine,Gieres}. Phase space does not necessarily represent initial value data or the state of a system at a fixed time. This interpretation is not covariant. More importantly, it assumes that an initial value problem exists and that the system occupies a well-defined state at any given moment. If the theory exhibits nonlocality in time, these assumptions may not be valid. 

Phase space has a symplectic structure. In \cite{Bernardes} the symplectic structure was argued to be
\begin{equation}
\Omega = \frac{1}{2}\omega\big(\delta\Phi,[Q_\Phi,\sigma]\delta\Phi\big),\label{eq:Omega2}
\end{equation}
where $\sigma$ is a commuting, grade 0 operator on $\H$ called the {\it sigmoid}. The sigmoid satisfies boundary conditions
\begin{equation}\lim_{t\to-\infty}\sigma = 0,\ \ \ \ \lim_{t\to\infty}\sigma = 1,\label{eq:sigma_boundary}\end{equation}
and is preserved through the BV inner product
\begin{equation}\omega(\sigma A,B) = \omega(A,\sigma B).\label{eq:sigma_omega}\end{equation}
This formula for symplectic structure appears to be largely new, but some precursors appear in \cite{Witten,Khavkine2,Lupo,Cho} and \cite{Benini} (remark 4.14). Because the sigmoid appears through a commutator, the symplectic form only has contribution from finite times where the sigmoid is not either 0 or 1. Therefore the sigmoid creates something analogous to a time slice for the symplectic structure. If the theory is local, the symplectic form can be localized to a Cauchy surface by choosing the sigmoid to act through multiplication by a unit step function which is zero in the past, and one in the future of the Cauchy surface. In \cite{Bernardes} it was argued that $\Omega$ has all the necessary properties to define a symplectic structure on phase space. The only property that was not established is nondegeneracy. We fill this gap in \cite{Bernardes5} with the (formal) construction of the Poisson bracket. When $\Omega$ is viewed as a 2-form on $\Ppre$, it is not a symplectic structure because it is degenerate along the gauge orbits. Sometimes it is called a {\it pre-symplectic form} for this reason. However we will refer to it simply as the symplectic form.  

Expanding the commutator $[Q_\Phi,\sigma]$ and assuming cyclicity, the symplectic structure appears to vanish on account of \eq{QdeltaPhi}. However, cyclicity breaks down because neither $\delta\Phi$ nor $\sigma$ vanish at temporal infinity. There are nonzero boundary terms. For the variational problem temporal boundaries are often not relevant, but on phase space they are critical in defining observables. 

\subsection{Symmetries and observables}

A vector field $\xi$ on $\Phat$ generates a transformation of $\Phi$ expressed through the Lie derivative
\begin{equation}\L_\xi \Phi = \xi_\Phi,\end{equation}
where $\xi_\Phi$ is a commuting element of $\H$ at grade 0 which may depend on $\Phi$. We refer to $\xi_\Phi$ as the {\it generating parameter} of the transformation. Taking the exterior derivative gives the {\it generating operator}, 
\begin{equation}\deltahat \xi_\Phi = \Xi_\Phi \deltahat\Phi,\label{eq:VPhi}\end{equation} 
which is a commuting operator on $\H$ at grade 0. The generating operator is assumed to be cyclic,
\begin{equation}\omega(\Xi_\Phi A,B) +\omega(A,\Xi_\Phi B) = 0.\end{equation}
Since this relates the action of $\Xi_\Phi$ at different grades, cyclicity can be taken to hold as a matter of definition without any boundary terms. We may describe $\xi_\Phi$ and $\Xi_\Phi$ in terms of products $\Xi_n$ so that
\begin{subequations}
\begin{align}
\xi_\Phi & = \sum_{n=0}^\infty \frac{1}{n!}\Xi_n(\underbrace{\Phi,\cdots,\Phi}_{n\text{ times}}),\\
\Xi_\Phi A & = \sum_{n=0}^\infty \frac{1}{n!}\Xi_{n+1}(\underbrace{\Phi,\cdots,\Phi}_{n\text{ times}},A).\label{eq:Xin}
\end{align}
\end{subequations}
Knowing \eq{deltaS} we can write down how the action changes under the flow of the vector field (up to temporal boundary contributions)
\begin{equation}
\L_\xi S = -\omega(\xi_\Phi,q_\Phi).\label{eq:LvS}
\end{equation}
We introduce the parameterized field $\Phi(s)$ and write this as
\begin{align}
\L_\xi S & = -\int_0^1 ds\, \frac{\d}{\d s}\omega(\xi_{\Phi(s)},q_{\Phi(s)})\nonumber\\
& = -\int_0^1 ds\left[\omega\left(\Xi_{\Phi(s)}\frac{\d \Phi(s)}{\d s},q_{\Phi(s)}\right)+\omega\left(\xi_{\Phi(s)},Q_{\Phi(s)}\frac{\d \Phi(s)}{\d s}\right)\right]\nonumber\\
& = \int_0^1 ds\left[\omega\left(\frac{\d \Phi(s)}{\d s},\Xi_{\Phi(s)} q_{\Phi(s)}\right)+\omega\left(Q_{\Phi(s)}\xi_{\Phi(s)},\frac{\d \Phi(s)}{\d s}\right)\right]\nonumber\\
& = -\int_0^1 ds\,\omega\left(\frac{\d \Phi(s)}{\d s},Q_{\Phi(s)}\xi_{\Phi(s)}-\Xi_{\Phi(s)} q_{\Phi(s)}\right).\label{eq:LvS2}
\end{align}
A {\it symmetry} of the action is defined by a generating parameter and generating operator which satisfy the condition
\begin{equation}Q_\Phi \xi_\Phi-\Xi_\Phi q_\Phi  = 0,\ \ \ \Phi\in \Phat.\label{eq:VE}\end{equation}
Gauge transformations automatically satisfy this condition. They are trivial symmetries because on phase space they generate no transformation at all. Therefore, symmetries which differ by gauge transformation should be seen as physically equivalent.

Noether's theorem states that every symmetry defines an observable, namely a conserved charge. Therefore it is helpful to characterize observables from the point of view of $L_\infty$ algebras. We start with a concept discussed for example in \cite{Erbin2} that we call an {\it off-shell observable}. This is a function on $\Phat$ which takes the form
\begin{equation}
F = -\int_0^1 ds\, \omega\left(\frac{\d \Phi(s)}{\d s},f_{\Phi(s)}\right),\ \ \ \Phi(s)\in\Phat,
\end{equation}
where $f_\Phi$ is an  anticommuting element of $\H$ at grade 1 which may depend on $\Phi$. We refer to $f_\Phi$ as the {\it characteristic state} of the observable. We also have a {\it characteristic operator} $F_\Phi$ which satisfies
\begin{equation}\deltahat f_\Phi = - F_\Phi\deltahat\Phi.\label{eq:deltaf}\end{equation}
This is an anticommuting operator on $\H$ at grade 1. We require that the characteristic operator is cyclic 
\begin{equation}\omega(F_\Phi A,B) +(-1)^{|A|}\omega(A,F_\Phi B) = 0,\label{eq:Fcyclic}\end{equation}
up to temporal boundary terms. The characteristic state and operator together must satisfy
\begin{equation} Q_\Phi f_\Phi + F_\Phi q_\Phi = 0,\ \ \ \Phi \in \Phat.\label{eq:Qf}\end{equation}
We may describe $f_\Phi$ and $F_\Phi$ in terms of products $F_n$ so that
\begin{subequations}
\begin{align}
f_\Phi & = \sum_{n=0}^\infty \frac{1}{n!}F_n(\underbrace{\Phi,\cdots,\Phi}_{n\text{ times}}),\\
F_\Phi A & = \sum_{n=0}^\infty \frac{1}{n!}F_{n+1}(\underbrace{\Phi,\cdots,\Phi}_{n\text{ times}},A).\label{eq:Fn}
\end{align}
\end{subequations}
The exterior derivative of the observable can be computed as with the action following \eq{deltaS2},
\begin{equation}\deltahat F = \omega(\deltahat\Phi, f_\Phi) \, . \label{eq:deltahatF}\end{equation}
This implies that the observable depends only on the value of $\Phi(s)$ at $s=1$. It is clear that the classical action itself is an off-shell observable. However, the action is special because it is gauge invariant. Usually off-shell observables are not identically gauge invariant. Gauge transformation instead gives 
\begin{align}
\L_\Lambda F & = -\omega(Q_\Phi\lambda_\Phi + \Lambda_\Phi q_\Phi, f_\Phi)\nonumber\\
& = -\omega(\lambda_\Phi,Q_\Phi f_\Phi)-\omega(q_\Phi,\Lambda_\Phi f_\Phi)\nonumber\\
& = \omega(\lambda_\Phi, F_\Phi q_\Phi)+\omega(\Lambda_\Phi f_\Phi,q_\Phi)\nonumber\\
& = \omega(F_\Phi\lambda_\Phi + \Lambda_\Phi f_\Phi,q_\Phi)\label{eq:LlambdaF}
\end{align}
up to temporal boundary terms. This is not zero. However, gauge invariance does hold {\it on-shell} with the caveat that the characteristic state $f_\Phi$ must be localized in time. This ensures that temporal boundary terms are absent, and that the observable does not diverge on-shell from integration over the infinite volume of time.\footnote{We can also worry about divergences from integration over the infinite volume of space. This issue however is somewhat different because often we are interested in finite energy solutions which are subject to falloff conditions towards spatial infinity. In this case the characteristic state may not need to be explicitly localized in spatial directions. Noether charges for example are not localized in spatial directions.}

This brings us to the notion of {\it on-shell observable}, 
\begin{equation}
F = -\int_0^1 ds\, \omega\left(\frac{\d \Phi(s)}{\d s},f_{\Phi(s)}\right),\ \ \ \Phi(s)\in\Ppre.\label{eq:Fonshell}
\end{equation}
This looks similar to an off-shell observable, but the characteristic state $f_\Phi$ is {\it a priori} defined only when $\Phi$ is a solution. It must also be localized in time to ensure that the observable is finite. The characteristic state satisfies 
\begin{equation}Q_\Phi f_\Phi = 0,\ \ \ \ \Phi\in \Ppre \, , \label{eq:Qf2} \end{equation}
which is the on-shell restriction of \eq{Qf}. The characteristic operator defined by 
\begin{equation}\delta f_\Phi = -F_\Phi\delta\Phi,\ \ \ \Phi\in\Ppre\end{equation} 
acts on tangent vectors to pre-phase space, and is required to be cyclic. The exterior derivative of an on-shell observable can be computed following \eq{deltaS2},
\begin{equation}\delta F =  \omega(\delta\Phi, f_\Phi),\end{equation}
which implies that the observable depends on $\Phi(s)$ only at $s=1$. With these conditions, the on-shell observable is a gauge invariant function of a classical solution, and is therefore a function on phase space $\P$. 

Let us make a comment. Computing an on-shell observable generally requires finding a path $\Phi(s)$ though the space of solutions from 0 to the desired target $\Phi$. However, the path can be allowed to leave the space of solutions if $f_\Phi$ can be extended off-shell in such a way that $F_\Phi$ remains cyclic. When this is possible, the observable will be somewhat easier to compute. We are able to achieve this simplification for Noether charges. However, the boundary observables recently constructed in \cite{Kaja} do not have a cyclic characteristic operator off-shell. This issue is important to keep in mind. 

Every off-shell observable defines an on-shell observable if it is finite after imposing the equations of motion.  However, an off-shell observable has additional structure. It defines a gauge invariant deformation of the classical action. If $F$ is an off-shell observable, the deformed action
\begin{equation}S' = S + \eps F\end{equation}
is invariant under the gauge transformation
\begin{equation}
\L_\Lambda \Phi = (Q_\Phi +\eps F_\Phi)\lambda_\Phi + \Lambda_\Phi(q_\Phi + \eps f_\Phi)
\end{equation}
up to first order in $\eps$. Another way of saying this is that an off-shell observable defines a tangent in the moduli space of cyclic $L_\infty$ structures. Some tangents are trivial, in the sense that they do not represent a physical deformation of the theory. These are the deformations generated by field redefinition, including those which shift the background. A field redefinition is implemented by a vector field on $\Phat$ with some generating parameter $\xi_\Phi$ and generating operator $\Xi_\Phi$. The change in the action under field redefinition can be seen from \eq{LvS2} to define an off-shell observable with the characteristic state
\begin{equation}f_\Phi = Q_\Phi \xi_\Phi - \Xi_\Phi q_\Phi.\label{eq:fred}\end{equation}
This can be shown to satisfy \eq{Fcyclic} and \eq{Qf}. This kind of observable carries no physical information because it vanishes on-shell, by virtue of the fact that it is a variation of the action. Off-shell observables whose difference is characterized by a field redefinition of the action should be seen as physically equivalent. Related discussion of gauge invariant deformations of the action appears in~\cite{Henneaux}. 

\subsection{Conserved charge}

A vector field $\xi$ on pre-phase space is a symmetry of the symplectic structure if
\begin{equation}\L_\xi\Omega = 0.\label{eq:LOmega}\end{equation}
The Lie derivative can be computed from Cartan's magic formula
\begin{equation}\L_\xi = \delta \iota_\xi +\iota_\xi\delta,\end{equation}
where $\iota_\xi$ denotes contraction with $\xi$. We assume that contraction is anticommuting and places the vector into the first entry of the differential form. Since the symplectic form is $\delta$-closed, \eq{LOmega} implies that $\iota_\xi\Omega$ is $\delta$-closed. If it is also $\delta$-exact, we learn that
\begin{equation}\iota_\xi\Omega + \delta F_\xi = 0,\label{eq:vF}\end{equation}
where $F_\xi$ is some function on pre-phase space. A vector field satisfying this equation is called a {\it Hamiltonian vector field}, and $F_\xi$ is the {\it conserved charge} associated to the vector field. As presented, a Hamiltonian vector field is a tangent to pre-phase space $\Ppre$. However, Hamiltonian vector fields can be considered equivalent if their difference generates a gauge transformation. In this way they are defined on phase space~$\P$. When discussing a theory defined by an $L_\infty$ algebra we are almost always working perturbatively, which also means that we are working locally near the vacuum $\Phi=0$ on pre-phase space. This means that for our purposes every $\delta$-closed 1-form is also $\delta$-exact, and every symmetry of the symplectic structure is generated by a Hamiltonian vector field with a conserved charge. We will not worry about whether the conserved charge can be defined globally on phase space in this paper. 

Now we state the main result: 
\begin{description}
\hypertarget{claim1}{\item{\bf Claim 1:}} If a vector field $\xi$ on $\Phat$ generates a symmetry of the action, its restriction to $\Ppre$ is a Hamiltonian vector field with conserved charge
\begin{equation}
F_\xi = -\int_0^1 ds\,\omega\left(\frac{\d \Phi(s)}{\d s}, [Q_{\Phi(s)},\sigma]\xi_{\Phi(s)}-[\Xi_{\Phi(s)},\sigma]q_{\Phi(s)}\right),\label{eq:Fv}
\end{equation} 
where $\sigma$ is the sigmoid satisfying \eq{sigma_boundary} and \eq{sigma_omega}, $\xi_\Phi$ and $\Xi_\Phi$ are the generating parameter and generating operator of the vector field $\xi$, and $\Phi(s)$ is a path through the space of fields subject to boundary conditions
\begin{equation}\Phi(0)= 0,\ \ \ \Phi(1)=\Phi\in\Ppre.\end{equation}
For intermediate $s$, the path $\Phi(s)$ can be off-shell. 
\end{description}
It is important that the Hamiltonian vector field originates from a symmetry of the action. If it is only a symmetry of the symplectic structure, the formula \eq{Fv} still applies except that $\Phi(s)$ must be taken on-shell \cite{Bernardes5}. The second term in the integrand \eq{Fv} drops out because it is proportional to the equations of motion. 

Let us give the proof. We start by computing the contraction
\begin{equation}
\iota_\xi \Omega = -\omega(\xi_\Phi,[Q_\Phi,\sigma]\delta\Phi),\label{eq:ixiomega}
\end{equation}
where we used $\iota_\xi\delta\Phi = \L_\xi\Phi = \xi_\Phi$. To proceed we need to open the commutator and use cyclicity. To do this while correctly accounting for temporal boundary terms, we use tau regularization~\cite{Bernardes}. The idea is to replace $\Phi\to\tau\Phi$ where $\tau$ is an operator that vanishes in the infinite past and future but is equal to 1 at finite times. Because the commutator $[Q_\Phi,\sigma]$ is nonzero only at finite time, this replacement should have no effect on the expression. But now $\tau\Phi$ vanishes in the infinite past and future, so we can use cyclicity without generating temporal boundary terms. To streamline notation we use the symbol $|_\tau$ to indicate that all instances of $\Phi$ and $\delta\Phi$ in an expression are replaced by $\tau\Phi$ and $\tau\delta\Phi$. For example,
\begin{equation}q_\Phi|_\tau = q_{\tau\Phi}.\end{equation}
Generally this is not zero even if $\Phi$ is a solution because $\tau\Phi$ is not a solution. However, $\tau\Phi$ is still a solution for finite times so $q_\Phi|_\tau$ will only be nonzero at temporal infinity. Applying tau regularization to \eq{ixiomega}, 
\begin{align}
\iota_\xi \Omega & = -\omega(\xi_{\Phi},[Q_{\Phi},\sigma]\delta\Phi)\big|_\tau\nonumber\\
& = \omega(Q_\Phi\xi_\Phi,\sigma \delta\Phi)\big|_\tau + \omega(\xi_\Phi,\sigma Q_\Phi\delta\Phi)\big|_\tau\nonumber\\
& = \omega(\Xi_\Phi q_\Phi,\sigma \delta\Phi)\big|_\tau-\omega(\xi_\Phi,\sigma\delta q_\Phi)\big|_\tau.
\label{eq:256}\end{align}
In the second step we opened the commutator and used cyclicity, and in the third we used \eq{VE} and \eq{deltaE}. Using cyclicity and graded antisymmetry of the BV inner product, 
\begin{align}
\iota_\xi\Omega & = -\omega(\Xi_\Phi\sigma \delta\Phi,q_\Phi)\big|_\tau - \omega(\sigma \xi_\Phi,\delta q_\Phi)\big|_\tau\nonumber\\
& = -\omega([\Xi_\Phi,\sigma] \delta\Phi,q_\Phi)\big|_\tau -\omega(\sigma \delta \xi_\Phi,q_\Phi) \big|_\tau- \omega(\sigma \xi_\Phi,\delta q_\Phi)\big|_\tau.
\end{align}
In the second step we commuted $\Xi_\Phi$ past the sigmoid and used \eq{VPhi}. Because the commutator with the sigmoid is localized to finite time we can drop the tau regulator. Then this term vanishes by the equations of motion. What remains is  $\delta$-exact,
\begin{equation}
\iota_\xi\Omega = \delta\Big(\omega(\sigma \xi_\Phi,q_\Phi)\big|_\tau\Big),
\end{equation}
which means that the conserved charge can be written
\begin{equation}F_\xi = -\omega(\sigma\xi_\Phi,q_\Phi)\big|_\tau.\label{eq:Fvtau}\end{equation}
We call this the {\it total derivative form} of the conserved charge. It appears to vanish by the equations of motion, but the tau regularization deforms the field off-shell near temporal infinity and gives a nonzero result. Therefore the conserved charge is written as a pure boundary term. This is similar to the symplectic structure, which is also formally zero and can be expressed as a pure boundary term \cite{Bernardes}. To obtain the conserved charge in a form which is localized in time, we parameterize the field as $\Phi(s)$ and express the charge as the integral of a total derivative with respect to $s$:
\begin{align}
F_\xi & = -\int_0^1 ds\,\frac{\d}{\d s}\omega(\sigma \xi_\Phi(s),q_\Phi(s))\big|_\tau\nonumber\\
& = -\int_0^1 ds\left[\left.\omega\left(\sigma \Xi_{\Phi(s)}\frac{\d \Phi(s)}{\d s},q_{\Phi(s)}\right)\right|_\tau+\left.\omega\left(\sigma \xi_{\Phi(s)},Q_{\Phi(s)}\frac{\d \Phi(s)}{\d s}\right)\right|_\tau\right]\nonumber\\
& = \int_0^1 ds\left.\omega\left(\frac{\d \Phi(s)}{\d s},\Xi_{\Phi(s)} \sigma q_{\Phi(s)}-Q_{\Phi(s)}\sigma \xi_{\Phi(s)}\right)\right|_\tau.\label{eq:253}
\end{align}
This is already localized to finite times on account of \eq{VE}. This however can be made more manifest by subtracting zero in the form
\begin{equation}\sigma\Big(\Xi_{\Phi(s)} q_{\Phi(s)}-Q_{\Phi(s)}\xi_{\Phi(s)}\Big)\Big|_\tau=0, \end{equation}
so that the sigmoid appears under a commutator. The tau regularization can then be dropped. This finishes the proof of \eq{Fv}.

The parameterized field $\Phi(s)$ allows us to localize the conserved charge explicitly using the universal algebraic objects we have introduced. However when applied to examples it is often tricky to eliminate the integration over $s$. We have found that it is usually easier to calculate the conserved charge in its total derivative form \eq{Fvtau} and integrate by parts to localize the integrand. If the sigmoid acts through multiplication by a function $\sigma(x)$, the general structure that appears from \eq{Fvtau} is 
\begin{equation}
F_\xi = -\int_M \mathrm{vol}\,\sigma (\nabla_\mu J^\mu)\big|_\tau,\label{eq:Fvtau_ex}
\end{equation}
where $J^\mu$ is the conserved current associated to the symmetry (evaluated on a tau regularized field). For finite times the current is evaluated on-shell and is conserved, so this expression only has contribution from temporal infinity. This is consistent with the fact that \eq{Fvtau} is proportional to the equations of motion. However now we can integrate by parts to find
\begin{equation}
F_\xi = \int_M \mathrm{vol}(\nabla_\mu\sigma)J^\mu.
\end{equation} 
Since the sigmoid is differentiated the conserved charge is localized to finite time, and the tau regulator can be dropped. In fact, this expression is exactly of the standard form of the conserved charge modified appropriately by the sigmoid. The total derivative form of the conserved charge \eq{Fvtau} is closely related to the idea, explained for example in Polchinski \cite{Polchinski}, of deriving conserved currents by varying the action with respect to a symmetry transformation with non-constant parameter. Comparing \eq{Fvtau} and \eq{LvS} it is clear that the sigmoid is playing the role of the non-constant parameter, which partly explains the general structure seen in \eq{Fvtau_ex}. We will demonstrate this in examples in section \ref{sec:examples}.

\subsection{Properties}  
\label{subsec:properites}
 
The conserved charge has five important properties:
\begin{description}
\item{\bf Gauge invariance:} The conserved charge is gauge invariant 
\begin{equation}\L_\Lambda F_\xi = 0.\end{equation} 
In particular it is a well-defined function on phase space $\P$. 
\item{\bf Linearity:} The conserved charge is a linear function of the vector field generating the symmetry. 
\item{\bf Noether's second theorem:} Any tangent $\Lambda$ to $\Phat$ which generates a gauge transformation has a vanishing conserved charge 
\begin{equation}F_\Lambda = 0. \end{equation} 
\item{\bf Localization:} The conserved charge is determined from data that is available at finite time. 
\item{\bf Conservation:} The conserved charge is independent of changes in the sigmoid which preserve the boundary conditions \eq{sigma_boundary}. 
\end{description}
Contracting \eq{vF} with the vector field of a gauge transformation we learn that
\begin{equation}\iota_\Lambda\iota_\xi\Omega + \mathcal{L}_\Lambda F_\xi = 0.\end{equation}
Gauge invariance follows because the vector field of a gauge transformation annihilates the symplectic structure \cite{Bernardes}. Linearity follows because the conserved charge depends linearly on the generating parameter and generating operator of the symmetry. The next property is sometimes referred to as Noether's second theorem \cite{Vitagliano}, though narrowly speaking the second theorem concerns the existence of Noether identities. To prove this statement we consider the total derivative form of the conserved charge \eq{Fvtau} and substitute the generating parameter of a gauge transformation
\begin{equation}
F_\Lambda = -\omega\big(Q_\Phi\lambda_\Phi + \Lambda_\Phi q_\Phi, \sigma q_\Phi\big)\big|_\tau.
\end{equation}
We use cyclicity and commute past the sigmoid to write this as
\begin{align}
F_\Lambda & = -\omega(\lambda_\Phi,[Q_\Phi,\sigma]q_\Phi)\big|_\tau-\omega(q_\Phi,[\Lambda_\Phi,\sigma]q_\Phi)\big|_\tau\nonumber\\
&\ \ \ -\omega(\lambda_\Phi,\sigma Q_\Phi q_\Phi)\big|_\tau-\omega(q_\Phi,\sigma\Lambda_\Phi q_\Phi)\big|_\tau.
\end{align}
The commutator terms are localized to finite time and vanish by the equations of motion. The third term vanishes on account of Noether identities \eq{QE}, while the fourth term is equal to its own negation using antisymmetry of $\omega$. Therefore $F_\Lambda=0$. Localization generalizes the familiar statement that the charge can be expressed as the integral of a local charge density over a spatial slice. In the presence of nonlocality this may no longer be possible, but still the charge is computable from data accessible at finite time, as determined by the sigmoid and the nature of the theory's nonlocality. Localization is manifest when the charge is written as \eq{Fv}. To prove conservation, we consider the difference of charges defined by sigmoids $\sigma$ and $\sigma'$. Using \eq{Fvtau} we have
\begin{equation}
F_\xi(\sigma) - F_\xi(\sigma') = -\omega\big((\sigma-\sigma')\xi_\Phi,q_\Phi\big)\big|_\tau.
\end{equation}
The difference of sigmoids is localized to finite time, which means we can drop the tau regularization. Then we find zero by the equations of motion.

The conserved charge is an observable with a characteristic state
\begin{equation}f_\Phi = [Q_\Phi,\sigma]\xi_\Phi-[\Xi_\Phi,\sigma]q_\Phi. \end{equation}
Assuming $\Phi$ is on-shell, we can confirm \eq{Qf2}:
\begin{align}
Q_\Phi f_\Phi & = Q_\Phi\Big([Q_\Phi,\sigma]\xi_\Phi-[\Xi_\Phi,\sigma]q_\Phi\Big)\nonumber\\
& = -[Q_\Phi,\sigma]Q_\Phi\xi_\Phi\nonumber\\
& = -[Q_\Phi,\sigma]\Xi_\Phi q_\Phi\nonumber\\
& = 0,\label{eq:Qfcons}
\end{align}
where we drop terms proportional to $q_\Phi$. We can also confirm that the characteristic operator is cyclic. We do this using the coalgebra formalism in appendix \ref{app:cyclicity}. However the characteristic state and operator do not satisfy \eq{Qf} off-shell. Therefore the conserved charge is only an on-shell observable. It does not define a gauge invariant deformation of the action.

\section{Coalgebra description of conserved charges} 
\label{sec:coalgebra}

We have taken some effort to express the conserved charge in terms of the equations of motion and generating parameter of the symmetry. For most important calculations this works well, but some questions require understanding algebraic relations between higher products. In this context it is often convenient to employ the coalgebra formalism (see \cite{Kajiura} and references therein). Below we give an overview of our results in this language. Our notation follows conventions used in literature on superstring field theory \cite{Erler5,Erler6}. See also \cite{Vosmera}. We describe the $L_\infty$ algebra as a nilpotent coderivation on the symmetric subspace of the tensor coalgebra~$T\H$. This is slightly different from the standard approach which uses a nilpotent coderivation on the symmetric coalgebra $S\H$. The difference comes down to certain factorial factors which we explain in appendix \ref{app:tensor}. The tensor coalgebra is more natural for discussion of cyclicity. 

The construction is based on the following cyclic coderivations on the tensor coalgebra $T\H$:
\begin{center}
\setlength{\tabcolsep}{10pt}
\renewcommand{\arraystretch}{1.5}
\begin{tabular}{|c|c|c|c|}
\hline
action & ${\bf M}$ & Grade 1, anticommuting & $\langle \omega|\pi_2\M =0 $\\
\hline
$\begin{matrix}\text{gauge} \\[-6pt] \text{transformations}\end{matrix}$ & $\llambda$ & Grade $-1$, anticommuting  & $\langle \omega|\pi_2\llambda =0 $ \\
\hline
symmetries & $\xxi$ & Grade 0, commuting & $\langle \omega|\pi_2\xxi =0 $ \\
\hline
observables & $\f$ & Grade 1, anticommuting & $\langle \omega|\pi_2\f =0 $\\
\hline
\end{tabular}
\end{center}
defining respectively the action, gauge transformations, symmetries, and observables. The final column expresses cyclicity. The coderivation $\M$ is formed from the products of the cyclic $L_\infty$ algebra of the theory. The classical action can be expressed
\begin{equation}
S = -\int_0^1 ds\,\omega\left(\frac{\d \Phi(s)}{\d s},\pi_1\M\frac{1}{1-\Phi(s)}\right),
\end{equation}
where 
\begin{equation}\frac{1}{1-\Phi} = 1_{T\H} +\Phi+\Phi\otimes\Phi+ \cdots\end{equation} 
is the group-like element of the tensor coalgebra generated from the field $\Phi$.  The $L_\infty$ relations can be stated as the condition that $\M$ is nilpotent: 
\begin{equation}\M^2 \, \mathrm{Sym} = 0,\label{eq:M2}\end{equation}
where $\mathrm{Sym}$ is the projector onto the symmetric part of the tensor coalgebra. The coderivation of a symmetry $\xxi$ must satisfy \cite{Erler7}
\begin{equation}
[\M,\xxi]\mathrm{Sym} = 0.\label{eq:Mv}
\end{equation}
Gauge transformations are symmetries of the form 
\begin{equation}
\xxi = [\M,\llambda],\label{eq:Mll}
\end{equation}
for a coderivation $\llambda$. Gauge transformations are automatically symmetries because $\M$ is nilpotent. The space of physically inequivalent symmetries can be identified with the cohomology of $\M$ in the space of cyclic coderivations at grade 0. This is sometimes called the {\it cyclic cohomology} of the $L_\infty$ algebra~\cite{Penkava,Moeller}. The coderivation $\f$ of an off-shell observable must satisfy
\begin{equation}[\M,\f]\mathrm{Sym} = 0.\end{equation}
This implies 
\begin{equation}\M' = \M + \eps \f\end{equation}
will be nilpotent to first order in $\eps$, and therefore represents a deformation of the cyclic $L_\infty$ structure. If 
\begin{equation}\f = [\M,\xxi],\end{equation}
where $\xxi$ is a coderivation at grade zero (not a symmetry), the deformation is the result of a field redefinition. The space of nontrivial deformations of the cyclic $L_\infty$ structure can be identified with the cohomology of $\M$ in the space of cyclic coderivations at grade~$1$, again the cyclic cohomology. 

The coderivations are related to the states and operators introduced earlier as follows:
\begin{center}
\renewcommand{\arraystretch}{1.5}
\begin{tabular}{|cc|cc|}
\hline
$\begin{matrix}\text{Euler-Lagrange} \\[-8pt] \text{state}\end{matrix}$ & \hspace{.8cm}$\displaystyle{q_\Phi = \pi_1{\bf M}\frac{1}{1-\Phi}}$\ \ \ \ \ \ & $\begin{matrix}\text{kinetic} \\[-8pt] \text{operator}\end{matrix}$  &\ \  $\displaystyle{Q_\Phi A = \pi_1\M \frac{1}{1-\Phi}A\frac{1}{1-\Phi} }$\\
\hline
$\begin{matrix}\text{gauge} \\[-8pt] \text{parameter}\end{matrix}$ & $\displaystyle{\lambda_\Phi = \pi_1\llambda \frac{1}{1-\Phi}}$ & $\begin{matrix}\text{gauge} \\[-8pt] \text{operator}\end{matrix}$  & $\displaystyle{\Lambda_\Phi A = \pi_1\llambda \frac{1}{1-\Phi}A\frac{1}{1-\Phi} }$\\
\hline
$\begin{matrix}\text{generating} \\[-8pt] \text{parameter}\end{matrix}$ & $\displaystyle{\xi_\Phi = \pi_1\xxi \frac{1}{1-\Phi}}$ & $\begin{matrix}\text{generating} \\[-8pt] \text{operator}\end{matrix}$  & $\displaystyle{\Xi_\Phi A = \pi_1\xxi\frac{1}{1-\Phi}A\frac{1}{1-\Phi} }$\\
\hline
$\begin{matrix}\text{characteristic} \\[-8pt] \text{state}\end{matrix}$ & $\displaystyle{f_\Phi = \pi_1\f \frac{1}{1-\Phi}}$ & $\begin{matrix}\text{characteristic} \\[-8pt] \text{operator}\end{matrix}$  & $\displaystyle{F_\Phi A = \pi_1\f\frac{1}{1-\Phi} A\frac{1}{1-\Phi} }$\\
\hline
\end{tabular}
\end{center}
Critical identities given earlier follow directly from the corresponding properties of the coderivations:
\begin{center}
\begin{tabular}{|c|c|}
\hline
& \\
Noether Identity & $\displaystyle{Q_\Phi q_\Phi =  \pi_1\M^2 \frac{1}{1-\Phi}=0}$ \\
&\\
\hline
&\\
Gauge transformation & $\displaystyle{\L_\Lambda \Phi = Q_\Phi \lambda_\Phi + \Lambda_\Phi q_\Phi = \pi_1[\M,\llambda]\frac{1}{1-\Phi}}$ \\
&\\
\hline
&\\
Symmetry & $\displaystyle{Q_\Phi \xi_\Phi-\Xi_\Phi q_\Phi  = \pi_1[\M,\xxi]\frac{1}{1-\Phi}=0}$    \\
&\\
\hline
&\\
Off-shell observable & $\displaystyle{Q_\Phi f_\Phi + F_\Phi q_\Phi = \pi_1[\M,\f]\frac{1}{1-\Phi}=0}$\\
&\\
\hline
\end{tabular}
\end{center}
Sometimes it is convenient to introduce higher products defined by acting the coderivations on group-like elements with two or more insertions. These appear, for example, in the generating operator for a gauge transformation,
\begin{equation}\Xi_\Phi A = \pi_1 [\M,\llambda]\frac{1}{1-\Phi}\otimes A\otimes \frac{1}{1-\Phi},\end{equation}
which may be rewritten as 
\begin{equation}\Xi_\Phi A = [Q_\Phi,\Lambda_\Phi] A+L_2^\Phi(\lambda_\Phi,A) + \Lambda^\Phi_2(q_\Phi,A),\label{eq:Xigauge}\end{equation}
after introducing
\begin{subequations}
\begin{align}
L_2^\Phi(A,B)& =\pi_1\M\left(\frac{1}{1-\Phi} A\frac{1}{1-\Phi}B\frac{1}{1-\Phi}+(-1)^{|A||B|}\frac{1}{1-\Phi}B\!\frac{1}{1-\Phi}A\frac{1}{1-\Phi}\right),\\
\Lambda_2^\Phi(A,B)& =\pi_1\llambda\left(\frac{1}{1-\Phi} A\frac{1}{1-\Phi}B\frac{1}{1-\Phi}+(-1)^{|A||B|}\frac{1}{1-\Phi} B\frac{1}{1-\Phi} A\frac{1}{1-\Phi}\right).
\end{align}
\end{subequations}
We can prove that gauge transformations satisfy the condition \eq{VE} defining a symmetry: 
\begin{align}
\Xi_\Phi q_\Phi - Q_\Phi \xi_\Phi & =\Xi_\Phi \left(\pi_1\M\frac{1}{1-\Phi}\right) - Q_\Phi \left(\pi_1[\M,\llambda]\frac{1}{1-\Phi}\right)\nonumber\\
& = \pi_1 [\M,\llambda] \left[\frac{1}{1-\Phi}\otimes \left(\pi_1\M\frac{1}{1-\Phi}\right)\otimes \frac{1}{1-\Phi}\right] \nonumber\\
& \ \ \ \ \ \ \ \ \ \ \ \ \ \ \ \ \ \ \ \ \ \ \  - \pi_1\M \left[\frac{1}{1-\Phi}\otimes\left(\pi_1[\M,\llambda]\frac{1}{1-\Phi}\right)\otimes \frac{1}{1-\Phi}\right]\nonumber\\
& = \pi_1\Big([\M,\llambda]\M - \M[\M,\llambda]\Big)\frac{1}{1-\Phi}\nonumber\\
& = -\pi_1[\M,[\M,\llambda]]\frac{1}{1-\Phi}=0.
\end{align}
This vanishes because \eq{Mll} satisfies \eq{Mv}. 

The conserved charge can be expressed 
\begin{equation}
F_\xi = -\int_0^1 ds\,\omega\!\left(\frac{\d\Phi(s)}{\d s}, \pi_1\Big(\M\sigma\xxi-\xxi\sigma\M\Big)\frac{1}{1-\Phi(s)}\right),\label{eq:Fvcoalg}
\end{equation}
where $\xxi$ is the coderivation representing the symmetry and $\sigma\M$ and $\sigma\xxi$ are coderivations defined by applying the sigmoid to the output of the products in $\M$ and $\xxi$. This expression corresponds to the final line of \eq{253}. The conserved charge is generated because the sigmoid prevents the commutation of $\xxi$ with $\M$. This can be expanded explicitly in terms of the products of the $L_\infty$ algebra. Integrating the parameter $s$ gives
\begin{equation}
F_\xi = -\sum_{n=0}^\infty\frac{1}{n+1}\omega(\Phi,f_n(\underbrace{\Phi,\cdots,\Phi}_{n\ \mathrm{times}})),
\end{equation}
where $f_n$ are the products defined by $\M\sigma\xxi-\xxi\sigma\M$. These are determined from the products $M_n$ and $\xi_n$ of the coderivations $\M$ and $\xxi$ as 
\begin{align}
f_n = \sum_{k=0}^n \Big[& M_{k+1}\Big(\sigma \xi_{n-k}\otimes \mathbb{I}^{\otimes k} + \cdots + \mathbb{I}^{\otimes k}\otimes \sigma \xi_{n-k}\Big)\nonumber\\
& -\xi_{k+1}\Big(\sigma M_{n-k}\otimes \mathbb{I}^{\otimes k} + \cdots + \mathbb{I}^{\otimes k}\otimes \sigma M_{n-k}\Big)\Big].\label{eq:fn}
\end{align}
The products $M_n$ and $\xi_n$ are graded symmetric, and acting on a string of $\Phi$s will produce $(k+1)$ times the same factor in the sum 
\begin{align}
f_n(\underbrace{\Phi,\cdots,\Phi}_{n\ \mathrm{times}}) = \sum_{k=0}^n (k+1)\Big[&M_{k+1}\big(\sigma \xi_{n-k}(\underbrace{\Phi,\cdots,\Phi}_{n-k\ \mathrm{times}}),\underbrace{\Phi,\cdots,\Phi}_{k\ \mathrm{times}}\big) \nonumber\\
&-\xi_{k+1}\big(\sigma M_{n-k}(\underbrace{\Phi,\cdots,\Phi}_{n-k\ \mathrm{times}}),\underbrace{\Phi,\cdots,\Phi}_{k\ \mathrm{times}}\big)\Big].\label{eq:fnPhi}
\end{align}
The products $M_n$ and $\xi_n$ are not canonically normalized. They would be if we were discussing $A_\infty$ algebras, but here we consider $L_\infty$ algebras, where the canonical normalization comes with an additional $1/n!$:
\begin{equation}
M_n=\frac{1}{n!}L_n,\ \ \ \ \ \ \xi_n = \frac{1}{n!}\Xi_n,\ \ \ \ \ \ f_n= \frac{1}{n!}F_n.\label{eq:SHTHnorm}
\end{equation}
On the right hand sides are the products introduced earlier in \eq{S}, \eq{Xin} and \eq{Fn}.
With this change of normalization \eq{fnPhi} is reexpressed
\begin{align}
F_n(\underbrace{\Phi,\cdots,\Phi}_{n\ \mathrm{times}}) = \sum_{k=0}^n \left(\begin{matrix}n \\ k\end{matrix}\right)\Big[& L_{k+1}\big(\sigma \Xi_{n-k}(\underbrace{\Phi,\cdots,\Phi}_{n-k\ \mathrm{times}}),\underbrace{\Phi,\cdots,\Phi}_{k\ \mathrm{times}}\big)\nonumber\\
&-\Xi_{k+1}\big(\sigma L_{n-k}(\underbrace{\Phi,\cdots,\Phi}_{n-k\ \mathrm{times}}),\underbrace{\Phi,\cdots,\Phi}_{k\ \mathrm{times}}\big)\Big].
\end{align}
From this we obtain the conserved charge as written in \eq{Fvintro} in the introduction.

We can also express the symplectic structure in the coalgebra formalism. A formula which sometimes comes up is 
\begin{align}
\Omega = \frac{1}{2}\omega(\delta\Phi,[Q,\sigma]\delta\Phi)+\int_0^1 ds\,\omega\bigg(\frac{\d\Phi(s)}{\d s},\pi_1\M\bigg[& \frac{1}{1-\Phi(s)}\sigma\delta\Phi(s)\frac{1}{1-\Phi(s)}\delta\Phi(s)\frac{1}{1-\Phi(s)}\nonumber\\
&\hspace{-1.5cm} -\frac{1}{1-\Phi(s)}\delta\Phi(s)\frac{1}{1-\Phi(s)}\sigma\delta\Phi(s)\frac{1}{1-\Phi(s)}
\bigg]\bigg).
\end{align}
This is naturally seen as a ``free'' symplectic structure which is corrected perturbatively by interactions.  Expanded in terms of products we get
\begin{equation}
\Omega = \frac{1}{2}\omega(\delta\Phi,[Q,\sigma]\delta\Phi)+\sum_{n=2}^\infty \frac{1}{(n-1)!}\omega\Big(\Phi,L_n\big(\sigma\delta\Phi,\delta\Phi,\underbrace{\Phi,\cdots,\Phi}_{n-2\text{\ times}}\big)\Big).
\end{equation}
When the sigmoid approaches $1$ in the future the pair of $\delta\Phi$s make an antisymmetric input for the products $L_n$, which are graded symmetric. In this way we see that the symplectic structure is localized to finite time. In \cite{Bernardes2} the symplectic structure of Witten's string field theory is written
\begin{equation}
\Omega = -\frac{1}{g^2}\left[\frac{1}{2}\big\langle \delta\Psi,[Q,\sigma]\delta\Psi\big\rangle +\big\langle\Psi,[\sigma\delta\Psi,\delta\Psi]\big\rangle\right],
\end{equation}
which is a version of this formula. The commutator in the second term is defined with the open string star product. The corresponding formula based on $SL(2,\mathbb{R})$ vertices was used in \cite{Bernardes4}.

\section{Examples}
\label{sec:examples}

In this section we apply the formalism to a few examples. It will not be necessary to discuss the cyclic $L_\infty$ algebra underlying the theory in each example, though in principle this can be done. Following \cite{Bernardes}, it is sufficient to employ DeWitt notation. The procedure behind this is explained in appendix~\ref{app:fermions}, but aside from a few signs, the structure is self-evident.

\subsection{Stress tensor of scalar field}

We consider the stress tensor of the scalar field $\phi$ in a potential $V(\phi)$
\begin{equation}
S = \int d^D x\left(-\frac{1}{2}\d_\mu\phi\d^\mu\phi -V(\phi)\right).
\end{equation}
For simplicity we work in flat space without spatial boundary. We expand the action in a fluctuation $\varphi$ around a background field $\phi$
\begin{equation}
S[\phi+\varphi]= S[\phi]-\int d^D x \varphi \big(-\Box\phi + V'(\phi)\big) -\frac{1}{2}\int d^D x\, \varphi\big(-\Box + V''(\phi)\big)\varphi + \cdots.
\end{equation}
This determines the Euler-Lagrange state
\begin{equation}
q(x) = -\Box \phi + V'(\phi).
\end{equation}
Translation symmetry is defined by the generating parameter
\begin{equation}
\L_v\phi = \xi(x) = v^\mu \d_\mu \phi,
\end{equation}
where $v^\mu$ represents a constant spacetime translation vector. We consider a sigmoid that acts through multiplication by a function $\sigma(x)$. The sigmoid matrix should take the form
\begin{equation}
\sigma(x,y) = \sigma(x)\delta^D(x-y).
\end{equation}
We evaluate the total derivative form of the conserved charge \eq{Fvtau}:
\begin{align}
F_v & = -\int d^D x d^D y\, \xi(x)\sigma(x,y)q(y)\big|_\tau\nonumber\\ 
& = -\int d^D x \,\sigma v^\mu\d_\mu \phi \big(-\Box\phi + V'(\phi)\big)\big|_\tau.
\end{align}
We integrate by parts a derivative off the $\Box$, and simplify the contribution from the potential
\begin{align}
F_v  = \int d^D x \big[-\d_\nu(\sigma v^\mu\d_\mu \phi)\d^\nu\phi -\sigma v^\mu\d_\mu V(\phi)\big]\big|_\tau.
\end{align}
Because of the tau regularization the scalar field vanishes at temporal infinity, and we can integrate by parts without generating boundary contributions. In the first term we evaluate the derivative and in the second we integrate by parts
\begin{align}
F_v & = \int d^D x \big[-(\d_\nu\sigma) v^\mu \d_\mu \phi \d^\nu\phi - \sigma v^\mu (\d_\mu\d_\nu\phi)\d^\nu\phi+(\d_\mu\sigma) v^\mu V(\phi)\big]\big|_\tau\nonumber\\
& =  \int d^D x \big[-(\d_\mu\sigma) v_\nu \d^\nu \phi \d^\mu\phi - \frac{1}{2}\sigma v^\mu \d_\mu (\d_\nu\phi\d^\nu\phi)+(\d_\mu\sigma) v^\mu V(\phi)\big]\big|_\tau.
\end{align} 
Integrating by parts in the middle term the sigmoid now always appears under a derivative, and the expression is localized to finite time. We can then drop the tau regularization to get
\begin{equation}
F_v = -\int d^D x (\d_\mu\sigma)v_\nu\left[\d^\mu\phi\d^\nu\phi -\eta^{\mu\nu}\left(\frac{1}{2}\d^\lambda\phi\d_\lambda\phi +V(\phi)\right)\right].\label{eq:scalar_Fxi}
\end{equation}
We recognize this in the form
\begin{equation}
F_v = -\int d^D x (\d_\mu\sigma)v_\nu T^{\mu\nu},
\end{equation}
where $T^{\mu\nu}$ is the stress tensor of the scalar field 
\begin{equation}
T^{\mu\nu} = \d^\mu\phi\d^\nu\phi -\eta^{\mu\nu}\left(\frac{1}{2}\d^\lambda\phi\d_\lambda\phi +V(\phi)\right).
\end{equation}
Note that the sign is correct. If the vector $v^\mu$ points to the future then $v_\mu$ points to the past, and the resulting Hamiltonian has positive kinetic energy.

For the sake of illustration we repeat this calculation using the localized form of the conserved charge \eq{Fv}. Here the sigmoid always appears under a derivative and we don't have to worry about integration by parts to arrange this. Instead, we have to integrate by parts to create a total derivative with respect to the auxiliary parameter $s$. In practice we have found that this is usually more difficult than computing the conserved charge in total derivative form. For the scalar field theory, the localized conserved charge \eq{Fv} takes the form 
\begin{equation}
F_v = \int_0^1 ds\int d^Dx\, \frac{\d \phi}{\d s}\Big([v^\mu\d_\mu,\sigma](-\Box\phi+V'(\phi))-[-\Box+V''(\phi),\sigma]v^\mu\d_\mu\phi\Big).
\end{equation}
The scalar field underneath the $s$ integral is implicitly parameterized by $s$. We evaluate the commutators to arrive at the expression
\begin{equation}
F_v = \int_0^1 ds\int d^Dx\, \left( \frac{\d \phi}{\d s}\Big[-(\d_\mu\sigma)v^\mu \Box +(\Box\sigma)v^\mu\d_\mu +2(\d_\mu\sigma)v_\nu \d^\mu\d^\nu\Big]\phi + (\d_\mu\sigma)v^\mu \frac{\d V(\phi)}{\d s}\right).
\end{equation}
The contribution from the potential is fairly easy to express as a total derivative with respect to $s$. The derivative contributions are more difficult to deal with. The key point is that the characteristic operator of the conserved charge is always cyclic. We prove this in appendix \ref{app:cyclicity}. What this means in the present context is that the operator in square brackets above is preserved if we integrate by parts to operate on the $s$ derivative factor. This takes a short time to check. It is useful to express the operator in a form which makes this manifest. This leads to an alternative expression
\begin{equation}
F_v = \int_0^1 ds\int d^Dx\, \left( \frac{\d \phi}{\d s}\Big[-\d^\nu (v^\mu\d_\mu\sigma) \d_\nu +\d^\mu (v^\nu\d_\mu\sigma)\d_\nu + \d_\nu (v^\nu\d_\mu\sigma)\d^\mu\Big]\phi + (\d_\mu\sigma)v^\mu \frac{\d V(\phi)}{\d s}\right).
\end{equation}
Except for the derivative in $\d_\mu\sigma$, the derivatives here act on everything that follows to the right. The first term in square brackets is mapped to itself upon integration by parts, and the second two terms are mapped into each other. Because of this we can factor out the derivative with respect to $s$ while multiplying by $1/2$:
\begin{align}
F_v & = \int_0^1 ds\int d^Dx\, \left( \frac{1}{2}\frac{\d}{\d s}\Big(\phi\Big[-\d^\nu (v^\mu\d_\mu\sigma) \d_\nu +\d^\mu (v^\nu\d_\mu\sigma)\d_\nu + \d_\nu (v^\nu\d_\mu\sigma)\d^\mu\Big]\phi\Big) + (\d_\mu\sigma)v^\mu \frac{\d V(\phi)}{\d s}\right).
\end{align}
Evaluating the integral over $s$,
\begin{align}
F_v & = \int d^Dx\, \left( \frac{1}{2}\phi\Big[-\d^\nu (v^\mu\d_\mu\sigma) \d_\nu +\d^\mu (v^\nu\d_\mu\sigma)\d_\nu + \d_\nu (v^\nu\d_\mu\sigma)\d^\mu\Big]\phi + (\d_\mu\sigma)v^\mu V(\phi) \right).
\end{align}
Next we integrate the leftmost derivative in square brackets by parts to obtain
\begin{equation}
F_v = \int d^Dx\, \left( \frac{1}{2}(v^\mu\d_\mu\sigma)\d^\nu\phi\d_\nu\phi - (v_\nu\d_\mu\sigma)\d^\mu\phi\d^\nu\phi + (v^\mu\d_\mu\sigma) V(\phi) \right).
\end{equation} 
From here we collect terms to arrive at the same expression \eq{scalar_Fxi}. Throughout this calculation the sigmoid always appeared under a derivative, so the charge is explicitly localized in time. 

\subsection{Hamiltonian of $p$-adic string theory}
 
Here we compute the Hamiltonian of $p$-adic string theory \cite{Brekke} and compare it to the derivative expansion obtained using the traditional Noether procedure by Moeller and Zwiebach \cite{Moeller2}. The relevant Lagrangian is nonlocal and its Hamiltonian is tricky to obtain by traditional methods. Aside from \cite{Moeller2}, at least one independent derivation of the Hamiltonian appears in \cite{Heredia} using the formalism of \cite{Llosa,Gomis,Gomis2}. The action is
\begin{align} \label{eq:1.158}
	S = {1\over g^2} \int dt \, \left(
	- {1 \over 2} \phi p^{- {1 \over 2} \Box } \phi + {1 \over p+1} \phi^{p+1}
	\right) \, ,
	\quad \quad
	p^{-{1 \over 2} \Box } = p^{{1 \over 2} {d^2 \over dt^2} },
\end{align}
where $\phi=\phi(t)$ is a real scalar field and $p>1$. For simplicity we work in $D=0+1$ dimensions and $t$ is the time coordinate. The kinetic operator $p^{- {1 \over 2} \Box }$ can be defined in terms of a derivative expansion
\begin{align} \label{eq:0.2}
	p^{- {1 \over 2} \Box } = \sum_{n=0}^\infty {1 \over n!} \left( {\ln p \over 2} {d^2 \over dt^2} \right)^n ,
\end{align}
or alternatively, through its representation as Weierstrass transform \cite{Moeller2}
\begin{align} \label{eq:1.160}
	p^{- {1 \over 2} \Box }\phi(t) 
	= {1 \over \sqrt{2 \pi \ln p}}\int_{-\infty}^\infty dt'  \, 
	\exp \left( - {(t-t')^2 \over 2 \ln p}\right) \phi(t').
\end{align}
By expanding the action in a fluctuation around a background field we can identify the Euler-Lagrange state
\begin{align}
	q(t) =  {1 \over g^2} \left(  p^{- {1 \over 2} \Box }\phi - \phi^p \right).
\end{align}
The theory has time translation symmetry defined by the generating parameter
\begin{align}
	\xi(t) = \dot{\phi},
\end{align}
where the dot denotes the time derivative. Finally, we choose the sigmoid to act through multiplication by $\sigma(t)$. The only substantial difference from the ordinary scalar field is the nonlocal kinetic operator appearing in the action.

We compute the total derivative form of the Hamiltonian \eq{Fvtau}
\begin{align} \label{eq:0.8}
	H = - {1 \over g^2} \int dt \; \sigma \dot{\phi}  \left(  p^{- {1 \over 2} \Box }\phi - \phi^p \right)  \bigg|_\tau.
\end{align}
We rewrite this expression as
\begin{align}
	H = - {1 \over g^2} \int dt \, \left[
	{1 \over 2} \left( [ p^{- {1 \over 2} \Box } ,\sigma] \dot{\phi} \right) \phi
	+ {1 \over 2} \sigma \left( p^{- {1 \over 2} \Box } \dot{\phi} \right) \phi
	+ {1 \over 2} \sigma \dot{\phi} \left( p^{- {1 \over 2} \Box }\phi \right) 
	- \sigma \dot{\phi} \phi^p
	\right] \bigg|_\tau.
\end{align}
We integrated $p^{- {1 \over 2} \Box } $ by parts acting on half of the first term in~\eqref{eq:0.8} and introduced the commutator with $\sigma$. The last three terms can be collected into
\begin{align}
	H = - {1 \over g^2} \int dt \, \left[
	{1 \over 2 }  \left( [ p^{- {1 \over 2} \Box } ,\sigma] \dot{\phi} \right) \phi 
	-  \sigma {d \over d t} \left(
	- {1 \over 2} \phi  p^{- {1 \over 2} \Box } \phi + {\phi^{p+1} \over p+1}
	\right) 
	\right]\bigg|_\tau.
\end{align}
After partially integrating the time derivative and moving $[ p^{- {1 \over 2} \Box } ,\sigma]$ to act on $\phi$, we obtain 
\begin{align} \label{eq:0.11}
	H = {1 \over g^2} \int_{-\infty}^\infty dt \, \left[
	 {1 \over 2 } \dot{\phi} [ p^{- {1 \over 2} \Box } ,\sigma] \phi 
	- \dot{\sigma} \left(
	-{1 \over 2} \phi  p^{- {1 \over 2} \Box } \phi + {\phi^{p+1} \over p+1}
	\right)
	\right] \, .
\end{align}
Each term is localized to finite time and the tau regulator is no longer necessary. This is the Hamiltonian for $p$-adic string theory. We checked that the localized form of the conserved charge leads to the same result.

The term in parenthesis in~\eqref{eq:0.11} is the Lagrangian for $p$-adic string theory~\eqref{eq:1.158}, and we observe a structural similarity to the traditional expression for Hamiltonian in terms of Legendre transformation. The first term should be the analogue of the $p\dot{q}$ contribution. This term can be expanded explicitly using the Weierstrass transform. Choosing the sigmoid as a unit step function centered at the origin we find
\begin{align}
	 \int_{-\infty}^\infty dt \,
	\dot{\phi} [ p^{- {1 \over 2} \Box } ,\theta] \phi 
	&=  {1\over \sqrt{2 \pi \ln p} }  \int_{-\infty}^0 dt \int_0^{\infty} dt'
	\exp \left( - {(t-t')^2 \over 2 \ln p}\right) 
	\left( \dot{\phi}(t) \phi(t') - \phi(t) \dot{\phi}(t') \right).
\end{align}
A related calculation appears in \cite{Bernardes}, where some care must be taken to discard the region of integration where the Weierstrass transform is not absolutely convergent. Evaluating the Lagrangian on-shell, the Hamiltonian reduces~to 
\begin{align} \label{eq:0.13}
	H = {1 \over 2g^2} {1 \over \sqrt{2 \pi \ln p}} \int_{-\infty}^0 dt \int_0^{\infty} dt'
	\exp \left( - {(t-t')^2 \over 2 \ln p}\right)  \left( \dot{\phi}(t) \phi(t') - \phi(t) \dot{\phi}(t') \right)
	+ {1 \over 2g^2}  {p-1 \over p+1 } \phi^{p+1}(0).
\end{align}
The Hamiltonian is not defined by the field and its derivatives on a time slice, even though we have chosen the sharpest possible sigmoid. However it is localized near $t=0$. Showing that the Hamiltonian is conserved and nonvanishing follows similar arguments discussed for the symplectic form in~\cite{Bernardes}. 

A different  expression for the $p$-adic Hamiltonian was derived in \cite{Moeller2} by expanding the action in derivatives and applying the Noether procedure. Introducing the notation
\begin{align}
	\phi_n \equiv {d^n \phi \over dt^n} (0)  \, ,
	\quad \quad 
	\text{for} \quad n = 0,1, 2, \cdots,
\end{align}
the authors found the Hamiltonian to be 
\begin{align} \label{eq:0.14}
	H =
	- {1 \over 2 g^2} \sum_{l=1}^\infty {1 \over l!} \left({1 \over 2} \ln p\right)^l
	\sum_{k=0}^{2l -1} (-1)^k \phi_{2l-k} \phi_k
	+  {1 \over 2 g^2 } {p-1 \over p+1} \phi_0^{p+1}.
\end{align}
Formally this is written in terms of derivatives of the field at $t=0$, but an infinite number of derivatives are needed. The on-shell Lagrangian term matches ours, so all we need to do is show that the double sum is same as the double integral in~\eqref{eq:0.13}. For this we perform a Taylor series expansion of $\phi(t)$ around $t=0$ inside the double integral and commute the integral/sums
\begin{align} \label{eq:0.16}
	 &{1 \over 2 g^2}{1 \over \sqrt{2 \pi \ln p} }   \int_{-\infty}^0 dt \int_0^{\infty} dt'
	\exp \left( - {(t-t')^2 \over 2 \ln p}\right)  \left( \dot{\phi}(t) \phi(t') - \phi(t) \dot{\phi}(t') \right)
	\nonumber \\ \nonumber
	&\hspace{1in}
	={1 \over 2 g^2}\sum_{n,m=0}^\infty {\phi_{n+1} \phi_m - \phi_n \phi_{m+1} \over n! m!} 
	 {1 \over \sqrt{2 \pi \ln p} }  \int_{-\infty}^0 dt \int_0^{\infty} dt'
	\exp \left( - {(t-t')^2 \over 2 \ln p}\right) t^n (t')^m 
	\\ 
	&\hspace{1in}=  {1 \over 2 g^2}\sum_{n,m=0}^\infty {\phi_{n+1} \phi_m - \phi_n \phi_{m+1} \over n! m!}
	\, \partial_\alpha^n \partial_\beta^m I(\alpha, \beta) \bigg|_{\alpha,\beta = 0},
\end{align}
where we introduce a generating function
\begin{align} \label{eq:0.17}
	I(\alpha,\beta) &= {1 \over \sqrt{2 \pi \ln p}}\int_{-\infty}^0 dt \int_0^\infty dt'  \exp \bigg(- {(t-t')^2 \over 2 \ln p} + \alpha t + \beta t' \bigg).
\end{align}
The integral is difficult to evaluate in closed form, but the complicated part actually cancels by the following symmetry argument. Note that the generating function satisfies
\begin{align} \label{eq:0.18}
	\partial_\alpha^n \partial_\beta^m I(\alpha, \beta) \bigg|_{\alpha,\beta = 0}
	= (-1)^{n+m} \partial_\alpha^m \partial_\beta^n I(\alpha, \beta) \bigg|_{\alpha,\beta = 0}.
\end{align}
This shows that terms with even $n+m$ cancel pairwise in the double sum~\eqref{eq:0.16} due to the multiplication of~\eqref{eq:0.18} with a combination of derivatives of $\phi$s that is antisymmetric under $n \leftrightarrow m$. Thus, only odd $n+m$ terms contribute. This allows us to replace $I(\alpha,\beta)$ with an antisymmetrized combination
\begin{align}
	I_{-} (\alpha, \beta) &= {1 \over 2} (I(\alpha,\beta) -I(-\alpha, -\beta)) 
	\\ \nonumber
	&= {1 \over \sqrt{2 \pi \ln p}}\int_{-\infty}^0 dt \int_0^\infty dt'  \exp \bigg(- {(t-t')^2 \over 2 \ln p} \bigg) \sinh(\alpha t + \beta t').
\end{align}
Therefore 
\begin{equation}{1 \over 2 g^2}\sum_{n,m=0}^\infty {\phi_{n+1} \phi_m - \phi_n \phi_{m+1} \over n! m!}
	\, \partial_\alpha^n \partial_\beta^m I(\alpha, \beta) \bigg|_{\alpha,\beta = 0} = {1 \over g^2} \sum_{\substack{n,m=0}}^\infty  {\phi_{n+1} \phi_{m} \over n! m!}
	\partial_\alpha^n \partial_\beta^m I_-(\alpha,\beta) \bigg|_{\alpha,\beta = 0}.\label{eq:Im_simp}
\end{equation}
Unlike~\eqref{eq:0.17}, the antisymmetric generating function $I_-(\alpha,\beta)$ can be evaluated in closed form using the simple substitution
\begin{align}
	u = t' - t \, ,\quad \quad v = t' + t,
\end{align}
which leads to
\begin{align}
	I_{-} (\alpha, \beta) &= 
	{1 \over \sqrt{2 \pi \ln p}} {1 \over 2}  \int_0^\infty du \exp\left(
	- {u^2 \over 2 \ln p}
	\right)
	\int_{-u}^u dv \sinh \left(
	{\beta - \alpha \over 2} u  + {\beta + \alpha \over 2} v
	\right)\nonumber\\
	&=  {1 \over \sqrt{2 \pi \ln p}}{1 \over 2}\int_0^\infty du \exp\left(
	- {u^2 \over 2 \ln p}
	\right) \left[ {2 \over \beta + \alpha}\cosh \left(
	{\beta - \alpha \over 2} u  + {\beta + \alpha \over 2} v
	\right) \right]_{v=-u}^{v=u}
	\nonumber\\
	&= {1 \over \sqrt{2 \pi \ln p}} {1 \over \beta + \alpha} \int_0^\infty du \exp\left(
	- {u^2 \over 2 \ln p}
	\right) \left( \cosh(\beta u) - \cosh(\alpha u) \right)
	\nonumber\\ 
	&= - {p^{\alpha^2/2} - p^{\beta^2/2} \over 2(\alpha + \beta)}.
\end{align}
To calculate the derivatives of $I_-(\alpha,\beta)$ we expand it around $\alpha,\beta=0$:
\begin{align} \label{eq:0.22}
	I_{-} (\alpha, \beta)  
	&=-{1 \over 2} \sum_{l=1}^\infty {1 \over l!} \left(
	{1 \over 2} \ln p
	\right)^l {\alpha^{2l} - \beta^{2l} \over \alpha + \beta}\nonumber
	\\
	&=-{1 \over 2} \sum_{l=1}^\infty {1 \over l!} \left(
	{1 \over 2} \ln p
	\right)^l  (\alpha - \beta) (\alpha^{2l - 2} + \alpha^{2l - 4} \beta^2 + \cdots
	+ \alpha^2 \beta^{2l-4} + \beta^{2l - 2})
	\nonumber \\
	&= -{1 \over 2} \sum_{l=1}^\infty {1 \over l!} \left(
	{1 \over 2} \ln p
	\right)^l   \sum_{k=0}^{2l-1} (-1)^k \alpha^{2l  - k -1} \beta^{k}.
\end{align}
Noting that
\begin{equation}
\sum_{m,n=0}^\infty\frac{\phi_{n+1}\phi_m}{n!m!}\d_\alpha^n\d_\beta^m\Big(\alpha^{2l  - k -1} \beta^{k}\Big)\Big|_{\alpha,\beta=0}=\phi_{2l-k}\phi_k,
\end{equation}
we can plug \eq{0.22} into \eq{Im_simp} to find
\begin{align}
	&{1 \over 2 g^2} {1 \over \sqrt{2 \pi \ln p} } \int_{-\infty}^0 dt \int_0^{\infty} dt' 
	\exp \left( - {(t-t')^2 \over 2 \ln p}\right)  \left( \dot{\phi}(t) \phi(t') - \phi(t) \dot{\phi}(t') \right) 
	\\ \nonumber
	 &\hspace{2.5in} = -
	{1 \over 2 g^2} \sum_{l=1}^\infty {1 \over l!} \left(
	{1 \over 2} \ln p
	\right)^l \sum_{k=0}^{2l - 1} (-1)^k  \phi_{2l-k} \phi_k.
\end{align}
This shows that we find the same Hamiltonian (and in extension, same energy for rolling tachyon
solutions) in p-adic string theory as Moeller and Zwiebach~\cite{Moeller2}.

\subsection{Surface charges}
\label{subsec:surface}

In the next two subsections we consider conserved charges in the presence of spatial boundaries. In this context an important class of observables are {\it surface charges}~\cite{Compere}---charges that originate from the boundary of a time slice. These arise as a residual effect of gauge transformations acting on spatial boundaries. Surface charges are in fact the only kind of conserved charge we expect to find in a quantum gravitational theory, because continuous global symmetries are not expected to exist  \cite{Banks,Harlow2}. Parallel discussion of surface charges in string field theory appears in \cite{Kaja}, and there is a close relation to our formalism.  

We begin our discussion with Yang-Mills theory,
\begin{equation}S = -\frac{1}{4g^2}\int_M \mathrm{vol}\, \Tr\big[F_{\mu\nu}F^{\mu\nu}\big],\end{equation}
where $g$ is the Yang-Mills coupling constant and the field strength is related to the Lie-algebra valued gauge field $A_\mu$ as
\begin{equation}F_{\mu\nu} = \d_\mu A_\nu - \d_\nu A_\mu - i[A_\mu,A_\nu].\end{equation}
We work in a curved spacetime $M$ with metric $g_{\mu\nu}$ and spatial boundary $\Gamma$. We use ``$\mathrm{vol}$'' to denote the canonical volume form on the manifold over which we are computing the integral. For $M$ the canonical volume form is $\mathrm{vol}= d^D x \sqrt{-\mathrm{det}(g_{\mu\nu})}$. The spatial boundary $\Gamma$ has an outward pointing unit normal form $n_\mu$ and a projection tensor 
\begin{equation}\gamma_\mu^\nu = \delta_\mu^\nu - n^\nu n_\mu.\end{equation}
The Euler-Lagrange state and generating parameter for the gauge symmetry are given by
\begin{align} \label{eq:0.25}
	q^\mu(x) = {1 \over g^2} \sqrt{-g}\mathcal{D}_\nu F^{\mu \nu} \, ,
	\quad \quad
	\xi_\mu (x) = \mathcal{D}_\mu \lambda,
\end{align}
where $\mathcal{D}_\mu = \nabla_\mu - i[A_\mu, \cdot]$ is spacetime+gauge covariant derivative and $\lambda$ is a Lie-algebra valued gauge parameter. The surface charge comes from  gauge transformations that act trivially on the spatial boundary \cite{Barnich}, which means that the generating parameter satisfies
\begin{equation}
\gamma_\mu^\alpha\xi_\alpha = 0\ \ \ \ (\mathrm{on\ }\Gamma).
\end{equation}
However, the gauge parameter $\lambda$ itself is generally not zero on $\Gamma$. More precisely, if we assume that the Yang-Mills equations hold without sources everywhere on the interior of $M$, it is necessary that the spatial boundary has more than one disconnected component (which may contain a singular point of the solution) on which $\lambda$ takes distinct values. Otherwise the surface charge will vanish. The total derivative form of the charge reads
\begin{align}
	F_\lambda = - {1 \over g^2} \int_M \mathrm{vol} \, \mathrm{Tr} \left[  \sigma (\mathcal{D}_\mu \lambda) ( \mathcal{D}_\nu F^{\mu \nu} )\right] \Big|_\tau.
\end{align}
We assume that the support of both $\sigma$ and $\tau$ extends all the way to the spatial boundary of $M$. Partially integrating the spacetime+gauge covariant derivative off of $\lambda$ inside the Lie algebra trace we find three terms:
\begin{align}
	F_\lambda = {1 \over g^2} \int_M \mathrm{vol} \, 
	\mathrm{Tr} \left[ (\nabla_\mu \sigma) \lambda ( \mathcal{D}_\nu F^{\mu \nu} )\right]\Big|_\tau
	&+  {1 \over g^2} \int_M \mathrm{vol} \, 
	\mathrm{Tr} \left[ \sigma \lambda ( \mathcal{D}_\mu \mathcal{D}_\nu F^{\mu \nu} )\right] \Big|_\tau
	\\ \nonumber
	&\hspace{1in}
	- {1 \over g^2} \int_M \mathrm{vol} \, \nabla_\mu \mathrm{Tr} \left[ \sigma \lambda ( \mathcal{D}_\nu F^{\mu \nu} )\right] \bigg|_\tau.
\end{align}
The first two terms do not contribute. The first is localized to finite time because the sigmoid appears under a derivative, and then it vanishes by the equations of motion $\mathcal{D}_\nu F^{\mu \nu} = 0$. The second is equal to zero by the identity
\begin{align} \label{eq:0.28}
	\mathcal{D}_\mu \mathcal{D}_\nu F^{\mu \nu} = {1 \over 2} [\mathcal{D}_\mu, \mathcal{D}_\nu] F^{\mu \nu} = - {i \over 2} [F_{\mu \nu}, F^{\mu \nu}] = 0.
\end{align}
Therefore the only nontrivial contribution comes from the spatial boundary $\Gamma$
\begin{align} \label{eq:0.29}
	F_\lambda = - {1 \over g^2} \int_\Gamma \mathrm{vol} \, n_\mu \mathrm{Tr} \left[ \sigma  \lambda ( \mathcal{D}_\nu F^{\mu \nu} )\right] \Big|_\tau.
\end{align}
Next we need to replace the spacetime covariant derivative $\nabla_\mu$ with the hypersurface covariant derivative $D_\mu$. We note that $n_\mu F^{\mu\nu}$ is a tensor on $\Gamma$ because
\begin{equation}\gamma^\nu_\alpha(n_\mu F^{\mu\alpha}) = n_\mu F^{\mu\nu}\ \ \ \ (\mathrm{on\ }\Gamma).\end{equation}
Therefore it is meaningful to compute its hypersurface covariant derivative:
\begin{align}
D_\nu(n_\mu F^{\mu\nu})&= \gamma_\nu^\alpha\nabla_\alpha(n_\mu F^{\mu\nu})\nonumber\\
& = (\gamma_\nu^\alpha\nabla_\alpha n_\mu)F^{\mu\nu} + n_\mu\gamma_\nu^\alpha\nabla_\alpha F^{\mu\nu}\nonumber\\
& = K_{\nu\mu}F^{\mu\nu} +n_\mu \nabla_\nu F^{\mu\nu} - n_\mu n_\nu n^\alpha\nabla_\alpha F^{\mu\nu}\ \ \ \ (\mathrm{on\ }\Gamma).
\end{align} 
The first term comes from substituting the extrinsic curvature tensor of $\Gamma$
\begin{equation}K_{\nu\mu} = \gamma_\nu^\alpha\nabla_\alpha n_\mu \ \ \ \ (\mathrm{on\ }\Gamma),\end{equation}
and the second two terms come from expanding the projection tensor. The first and last term drop out because the field strength is contracted with symmetric objects, leaving
\begin{equation}
\label{eq:0.30a}
D_\nu(n_\mu F^{\mu\nu}) = n_\mu \nabla_\nu F^{\mu\nu} \ \ \ \ (\mathrm{on\ }\Gamma).
\end{equation}
With this we can replace the spacetime+gauge covariant derivative in \eq{0.29} with the hypersurface+gauge covariant derivative, which we can integrate by parts on the boundary inside the trace. The result is 
\begin{align} \label{eq:0.30}
	F_\lambda = {1 \over g^2} \int_\Gamma \mathrm{vol} \, n_\mu \mathrm{Tr} \left[
	\sigma (\mathcal{D}_\nu \lambda) F^{\mu \nu}
	\right] \Big|_\tau
	  +{1 \over g^2} \int_\Gamma \mathrm{vol} \, n_\mu \mathrm{Tr} \left[
	(D_\nu \sigma) \lambda F^{\mu \nu}\right]\Big|_\tau,
\end{align}
The tau regulator ensures that integration by parts does not generate further contributions from the boundary of $\Gamma$. The first term should drop out because the projection of the generating parameter $\xi_\mu = \mathcal{D}_\mu \lambda$ onto the spatial boundary vanishes.  However we need to be careful about this because of the tau regulator. The generating parameter vanishes on the boundary for finite times, but as the gauge field eventually transitions to zero in the infinite future $\mathcal{D}_\mu \lambda$ might become nonzero. This creates what we will call an {\it infinity corner term}. The infinity corner term is unphysical because it depends in detail on the tau regulator. We will assume that the tau regularization is chosen consistently with the boundary conditions in such a way that the infinity corner term vanishes. This leaves the second term in \eq{0.30},
\begin{align} \label{eq:0.34}
	F_\lambda = - {1 \over g^2} \int_\Gamma \mathrm{vol} \, (D_\mu \sigma) n_\nu  \mathrm{Tr} \left[
	\lambda F^{\mu \nu} \right].
\end{align}
The tau regulator can be removed because the sigmoid appears under a derivative. This is the known expression for the surface charge in Yang-Mills theory \cite{Abbott}, generalized to include the sigmoid. It computes the non-abelian gauge charge that serves to source the field configuration.

We describe one prescription which eliminates the infinity corner term. The tau regularization can operate as  
\begin{equation}\tau A_\mu = \theta(T-|t|) A_\mu^* + \theta(|t|-T)A_\mu',\end{equation}
where $T>0$ is the characteristic time of the tau regulator which is taken to be very large, and $\theta$ is the Heaviside step function. For $|t|<T$ the gauge field is equated with a solution $A_\mu^*$ representing a point in phase space. For $|t|>T$ the gauge field is equated with an off-shell field configuration $A_\mu'$ which satisfies the Neumann boundary condition,
\begin{equation}n_\mu F^{\mu\nu}=0\ \ \ \ \text{(on }\Gamma),\end{equation}
and is taken to vanish in the distant past and future. The infinity corner term in~\eqref{eq:0.30} is absent provided that $A_\mu^*$ matches continuously with $A_\mu'$ at $t=T$ on the boundary. Due to the Neumann boundary condition, it is not necessary to make any particular assumption about $\lambda$ for $t>T$. 

Following a similar procedure we can derive the surface charge in general relativity. The Einstein-Hilbert action is   
\begin{equation}S = \frac{1}{2\kappa}\int_M \mathrm{vol}\,R,\end{equation}
where $\kappa = 8\pi G$ and $R$ is the Ricci scalar. In principle the action also requires a GHY boundary term \cite{York,Gibbons}, but we do not need to discuss it. The Euler-Lagrange state and the generating parameter for diffeomorphisms takes the form
\begin{align}
	q^{\mu \nu}(x) = - {1 \over 2\kappa} \sqrt{-g} \, G^{\mu \nu},
	\quad \quad
	\xi_{\mu \nu}(x) = \nabla_{\mu} v_{\nu}+ \nabla_{\nu} v_{\mu},
\end{align}
where $G_{\mu \nu}$ is the Einstein tensor and $v^\mu$ is a vector field. To define a surface charge we assume that the diffeomorphism preserves the spatial boundary, 
\begin{equation}
\gamma_\mu^\alpha v_\alpha = v_\mu \ \ \ \ (\mathrm{on\ }\Gamma),\label{eq:tangent2}
\end{equation} 
and the boundary metric 
\begin{equation}
\gamma_\mu^\alpha\gamma_\nu^\beta \xi_{\alpha\beta} = 0 \ \ \ \ (\mathrm{on\ }\Gamma).\label{eq:xivan}
\end{equation}
This is equivalent to saying that $v^\mu$ is an isometry of the boundary metric,
\begin{equation}
D_{\mu}v_{\nu}+D_{\nu}v_{\mu}=0 \ \ \ \ (\mathrm{on\ }\Gamma).\label{eq:0.37b}
\end{equation}
If the vacuum Einstein equation holds everywhere on the interior of $M$, the surface charge can be nonzero only if $M$ has more than one boundary component where the vector field $v^\mu$ takes distinct values. The total derivative form of the conserved charge reads 
\begin{align} \label{eq:0.37a}
	F_v =  {1 \over \kappa} \int_M \mathrm{vol} \, \sigma (\nabla_{\mu} v_{\nu}) G^{\mu \nu} \big|_\tau.
\end{align}
We have to explain the meaning of the tau regulator here. We assume that the tau regulator deforms the metric $g_{\mu\nu}$ to a nearby reference metric $g^*_{\mu\nu}$ in the infinite past and future. Under suitable conditions, the nearby metric could be flat space $\eta_{\mu\nu}$. We also assume that the diffeomorphism $v$ is deformed into a reference diffeomorphism $v^*_\mu$ which satisfies both \eq{tangent2} and \eq{xivan} from the point of view of the reference metric. We partially integrate the derivative off of $v_\nu$ to obtain three terms
\begin{align}
	F_v = -{1 \over \kappa}  \int_M \mathrm{vol} \, (\nabla_{\mu} \sigma) v_{\nu} G^{\mu \nu}\big|_\tau
	- {1 \over \kappa} \int_M \mathrm{vol} \, \sigma v_{\nu} \nabla_{\mu} G^{\mu \nu} \big|_\tau
	+ {1 \over \kappa} \int_M \mathrm{vol} \, \nabla_{\mu} \left( \sigma v_{\nu} G^{\mu \nu} \right) \big|_\tau.
\end{align}
The bulk terms vanish as a consequence of equation of motion $G_{\mu \nu} = 0$ (after lifting tau regulator) and conservation of the Einstein tensor $\nabla_\nu G^{\mu \nu} = 0$. We are therefore left with
\begin{align} \label{eq:0.37}
	F_v =  {1 \over \kappa}  \int_\Gamma \mathrm{vol} \, n_{\mu} \sigma v_{\nu} G^{\mu \nu} \big|_\tau.
\end{align}
This shows that the charge is localized to the spatial boundary. To show that it is localized in time as well, we use the Gauss-Codazzi relation (see e.g. equation 10.2.24 in~\cite{Wald}) 
\begin{align} \label{eq:0.39}
	G^{\mu \nu} n_{\mu} \gamma_{\nu}^{\beta} = - \kappa D_\alpha T^{\alpha \beta}  \ \ \ \ (\mathrm{on\ }\Gamma),
\end{align}
where $T^{\alpha \beta}$ is the~\emph{Brown-York stress tensor}~\cite{Brown}
\begin{align} \label{eq:0.41}
	T^{\alpha \beta} =- {1 \over \kappa} \left( K^{\alpha \beta} - K \gamma^{\alpha \beta} \right)  \ \ \ \ (\mathrm{on\ }\Gamma).
\end{align}
The Brown-York stress tensor is conserved on the boundary when the metric is on shell, as can be seen from \eqref{eq:0.39}. The Gauss-Codazzi relation therefore allows us to rewrite the charge as 
\begin{align}
	F_v = -\int_\Gamma \mathrm{vol} \, \sigma v_{\beta} D_{\alpha} T^{\alpha \beta} \big|_\tau,
\end{align}
after using \eq{tangent2}. Partially integrating the boundary covariant derivative,
\begin{align} \label{eq:0.42}
	F_v = -\int_{\partial \Gamma} \mathrm{vol} \, \sigma \tau_\alpha  v_\beta^* (T^{\alpha \beta})^* 
	+ \int_\Gamma \mathrm{vol} \,  \sigma (D_{\alpha} v_{\beta}) T^{\alpha \beta}\big|_\tau 
	+ \int_\Gamma \mathrm{vol} \, (D_{\alpha} \sigma) v_{\beta} T^{\alpha \beta}\big|_\tau.
\end{align}
In the first term all of the objects are evaluated at temporal infinity, where due to the tau regulator we have the reference metric $g_{\mu\nu}^*$ and diffeomorphism $v_\mu^*$. We define $\tau_\mu$ as the future pointing unit normal form at temporal infinity on $\Gamma$. The second contribution is supposed to be zero because the generating parameter of the diffeomorphism vanishes when projected on the boundary, \eq{xivan}. However, the tau regularization could generate an infinity corner term. As before, we assume that the tau regularization is chosen so that this term vanishes. Finally, in the last contribution the sigmoid appears under a derivative, so we can drop the tau regulator. We find
\begin{align}
	F_v = -\int_{\partial \Gamma} \mathrm{vol} \, \sigma \tau_\alpha  v_\beta^* (T^{\alpha \beta})^* 
	+ \int_\Gamma \mathrm{vol} \, (D_{\alpha} \sigma) v_{\beta} T^{\alpha \beta}.
\end{align}
A moment's thought reveals that the first term can be written in the same way as the second by replacing the time slice at infinity with the derivative of the sigmoid. This gives 
\begin{align} 
	F_v = 
	 \int_\Gamma \mathrm{vol} \, (D_{\alpha} \sigma)\Big[v_{\beta} T^{\alpha \beta} - v_\beta^* (T^{\alpha \beta})^* \Big].
\end{align}
This is the well-known formula for the diffeomorphism charge in general relativity \cite{Brown}, modified appropriately by the sigmoid and with a subtraction from the reference background. Note that we
did not assume any particular spatial boundary condition on the metric or make use of the GHY boundary term to obtain this result.

In asymptotically flat spacetimes we can take Minkowski space as the reference background, and the subtraction is necessary to define a finite surface charge. In $D$ dimensions the volume of the sphere at spatial infinity increases as $r^{D-2}$ for large radial coordinate~$r$, while the extrinsic curvature of the sphere decreases only as $1/r$. The leading divergence at large $r$ is removed by subtraction from flat space. The subleading contribution is finite if the metric approaches Minkowski as $1/r^{D-3}$. An important check on our calculation for asymptotically flat spacetimes is that the second term in \eq{0.42} vanishes. This term represents the flux of gravitational mass/momentum out of the spatial boundary, and must be zero otherwise the surface charge is not conserved. If the metric approaches Minkowski as $1/r^{D-3}$ the Christoffel symbols fall off as $1/r^{D-2}$, which implies that the right hand side of the boundary Killing equation \eq{0.37b} will also fall off as $1/r^{D-2}$. The Killing equation then cancels the volume of the sphere in the second term of \eq{0.42}, which will then vanish as $1/r$ from the extrinsic curvature. Incidentally there is no infinity corner term, as the tau regulator can easily deform to Minkowski space consistently with the asymptotically flat falloff condition on the metric.

\subsection{Yang-Mills Killing charge}

As a final example we consider the Killing charge of Yang-Mills theory in the presence of a spatial boundary. This is not a surface charge, but nevertheless there is a nontrivial contribution from the spatial boundary. 

We consider spacetime $M$ with an isometry represented by Killing vector $v^\mu$
\begin{equation}
\nabla_{\mu}v_{\nu} +\nabla_{\nu}v_{\mu}= 0.
\end{equation}
We assume that the Killing vector is tangent to the boundary
\begin{equation}\gamma_\mu^\alpha v_\alpha = v_\mu \ \ \ \ (\mathrm{on\ }\Gamma).\end{equation}
Translating along the Killing vector changes the gauge field as
 \begin{align} \label{eq:0.49}
 	\xi_\mu(x) = \mathcal{L}_v A_\mu &= v^\nu ( \nabla_\nu A_\mu )  + (\nabla_{\mu} v^\nu) A_{\nu} \\
 	&= v^\nu F_{\nu \mu} + \mathcal{D}_\mu (v^\nu A_\nu) \nonumber.
 \end{align}
The expression in the second line is helpful for evaluating the conserved charge. We assume that the gauge field takes a fixed value on the boundary which does not break the Killing symmetry of the background spacetime. This means that the projection of the generating parameter on the boundary should vanish:
\begin{equation}\gamma_\mu^\nu\xi_\nu = 0  \ \ \ \ (\mathrm{on\ }\Gamma). \label{eq:0.49b}\end{equation}
Taking the Euler-Lagrange state from \eq{0.25}, the total derivative form of the conserved charge is 
\begin{align} \label{eq:0.50}
	F_v = - {1 \over g^2} \int_M \mathrm{vol} \,  \mathrm{Tr} \big[ \sigma 
	\left( v^\nu F_{\nu \mu} + \mathcal{D}_\mu (v^\nu A_\nu) \right) 
	\mathcal{D}_\rho F^{\mu \rho} \big] \big|_\tau.
\end{align}
To see how to simplify it is helpful to recall the form of the stress tensor in Yang-Mills theory:
\begin{align} \label{eq:0.51}  
	T^{\mu \nu} = \mathrm{Tr} \left[
	F^{\mu \alpha}  F^{\nu}_{\; \; \alpha}
	-{1 \over 4} g^{\mu \nu} F_{\alpha \beta} F^{\alpha \beta} 
	\right].
\end{align}
This is symmetric and gauge invariant. The divergence of the stress tensor is
\begin{equation}
	\nabla_\mu T^{\mu \nu} = \mathrm{Tr} \left[
	(\mathcal{D}_\mu F^{\mu \alpha})  F^{\nu}_{\; \; \alpha}
	+ F^{\mu \alpha} (\mathcal{D}_\mu F^{\nu}_{\; \; \alpha})
	-{1 \over 2} F^{\alpha \beta} (\mathcal{D}^\nu  F_{\alpha \beta} )
	\right].
\end{equation}
To simplify we contract the Bianchi identity,
\begin{align}
	\mathcal{D}_\mu F_{\alpha \beta} + \mathcal{D}_\alpha F_{\beta \mu} + \mathcal{D}_\beta F_{\mu \alpha} = 0,
\end{align}
with $F^{\alpha\beta}$ to find 
\begin{align}
	F^{\alpha \beta} \mathcal{D}_\mu F_{\alpha \beta} = 2 F^{\alpha\beta} \mathcal{D}_\alpha F_{\mu \beta}.
\end{align}
The divergence of the stress tensor then reduces to 
\begin{align}
	\nabla_\mu T^{\mu \nu}= \mathrm{Tr} \left[
	(\mathcal{D}_\mu F^{\mu \alpha})  F^{\nu}_{\; \; \alpha}
	\right].
\end{align}
This allows us to write the first term in \eq{0.50} using the divergence of the Yang-Mills stress tensor. Meanwhile, the second term can be written as a total derivative using \eq{0.28}
\begin{equation}\mathrm{Tr} \left[ \left( \mathcal{D}_\mu (v^\nu A_\nu) \right)
	\mathcal{D}_\rho F^{\mu \rho} \right] = \nabla_\mu\mathrm{Tr} \left[v^\nu A_\nu 
	\mathcal{D}_\rho F^{\mu \rho} \right].\end{equation}
Therefore we obtain
\begin{align} 
	F_v &= {1 \over g^2}  \int_M \mathrm{vol} \, \sigma \, v_\nu \nabla_\mu T^{\mu \nu}\big|_\tau
	- {1 \over g^2} \int_M \mathrm{vol} \, \sigma \, \nabla_\mu\mathrm{Tr} \left[v^\nu A_\nu 
	\mathcal{D}_\rho F^{\mu \rho} \right]
	\big|_\tau.
\end{align}
Partially integrating the covariant derivatives,
\begin{align} \label{eq:0.56}
	F_v = &- {1 \over g^2}  \int_M \mathrm{vol} \, (\nabla_\mu \sigma)  v_\nu T^{\mu \nu}\big|_\tau
	+{1 \over g^2}  \int_{\Gamma} \mathrm{vol} \, \sigma  n_\mu v_\nu T^{\mu \nu} \big|_\tau
	\\ \nonumber
	&+ {1 \over g^2} \int_M \mathrm{vol} \,  (\nabla_\mu \sigma) \, \mathrm{Tr} \left[  v^\nu A_\nu
	\mathcal{D}_\rho F^{\mu \rho}  \right]\big|_\tau
	- {1 \over g^2} \int_\Gamma \mathrm{vol} \, \sigma  n_\mu \mathrm{Tr} \left[  v^\nu A_\nu
	\mathcal{D}_\rho F^{\mu \rho}  \right]
	\big|_\tau.
\end{align}
In the first and third terms we can drop the tau regularization. The third term then vanishes by equations of motion. This leaves the expected expression for the Killing charge in terms of the energy momentum tensor, but there is remaining contribution from the spatial boundary:
\begin{align} \label{eq:0.58a}
	F_v = &- {1 \over g^2}  \int_M \mathrm{vol} \, (\nabla_\mu \sigma) v_\nu T^{\mu \nu}
	+{1 \over g^2}  \int_{\Gamma} \mathrm{vol} \, \sigma  n_\mu \Big(v_\nu T^{\mu \nu} 
	-  \mathrm{Tr} \left[ v^\nu  A_\nu
	\mathcal{D}_\rho F^{\mu \rho}  \right]\Big)
	\Big|_\tau.
\end{align}
Consider the term 
\begin{align} \label{eq:0.57}
	n_\mu v_\nu T^{\mu \nu}
	= n_\mu \mathrm{Tr} \left[ F^{\mu \rho} (v^\nu F_{\nu\rho})\right]  \ \ \ \ (\mathrm{on\ }\Gamma).
\end{align} 
The factor in parentheses can be written in terms of the generating parameter \eq{0.49}
\begin{align}
	n_\mu v_\nu T^{\mu \nu}
	= n_\mu \mathrm{Tr} \left[ F^{\mu \rho} \big(\xi_\rho-\mathcal{D}_\rho(v^\nu A_\nu)\big)\right]  \ \ \ \ (\mathrm{on\ }\Gamma).
\end{align}
Because the expression is contracted with the unit normal, antisymmetry of the field strength implies that we can project the factor in parentheses onto the spatial boundary. The generating parameter then drops out on account of the boundary condition \eq{0.49b} excepting perhaps an infinity corner term, which as in previous examples we assume vanishes after an appropriate choice of tau regularization. Finally we have 
\begin{align}
	n_\mu v_\nu T^{\mu \nu}
	= -n_\mu \mathrm{Tr} \left[ F^{\mu \rho} \mathcal{D}_\rho(v^\nu A_\nu)\right]  \ \ \ \ (\mathrm{on\ }\Gamma),
\end{align}
which can be added to the second term in \eq{0.58a} to form a total derivative
\begin{align}
	F_v = &- {1 \over g^2}  \int_M \mathrm{vol} \, (\nabla_\mu \sigma) v_\nu T^{\mu \nu}
	-{1 \over g^2}  \int_{\Gamma} \mathrm{vol} \, \sigma  n_\mu 
	\mathrm{Tr} \left[ \mathcal{D}_\rho(v^\nu  A_\nu
	 F^{\mu \rho})  \right]
	\big|_\tau.
\end{align} 
The spacetime+gauge covariant derivative can be replaced by the spacetime covariant derivative because of cyclicity of the trace. This in turn can be replaced by the hypersurface covariant derivative on account of \eq{0.30a}. Therefore 
\begin{align}
	F_v = &- {1 \over g^2}  \int_M \mathrm{vol} \, (\nabla_\mu \sigma) v_\nu T^{\mu \nu}
	-{1 \over g^2}  \int_{\Gamma} \mathrm{vol} \, \sigma  
	D_\rho \Big(\mathrm{Tr} \left[ v^\nu  A_\nu
	 n_\mu F^{\mu \rho}\right]\Big)
	\Big|_\tau.
\end{align}
Integrating by parts on the boundary the sigmoid appears under a derivative. We can drop the tau regulator to find the result 
\begin{align} \label{eq:0.61}
	F_v =  - {1 \over g^2}  \int_M \mathrm{vol} \, (\nabla_\mu \sigma)  v_\nu T^{\mu \nu}
	&- {1 \over g^2} \int_\Gamma \mathrm{vol} \, (D_\mu  \sigma)  n_\nu v_\rho \mathrm{Tr} \left[ F^{\mu \nu} A^\rho \right].
\end{align}
The result can be compared to appendix A of \cite{Harlow3}.

\section{Discussion}

In this paper we constructed conserved charges associated to symmetries of the action in the language of $L_\infty$ algebras. The result can be viewed as a generalization of Noether's theorem. This generalization is  relevant especially for higher derivative and nonlocal field theories where the traditional construction of conserved charges may be difficult to apply. 

At least in some cases, our formalism gives the correct conserved charges in the presence of spatial boundaries with minimal input from the variational principle. This is encouraging because in string theory we do not know how to formulate the variational principle. The standard recipe is to express the spacetime action in terms of a Lagrangian density which depends on at most first derivatives of the field. However, this is not possible in string field theory because the interactions involve an infinite number of derivatives. However, one can formulate the variational problem for the free string action in the presence of boundaries \cite{Stettinger,Firat,Maccaferri1,Maccaferri2,Gen}. In principle, background independence \cite{Sen2,Sen3} should then define the variational problem for the fully interacting theory. But it is difficult to make this idea concrete. A further complication is that the action requires an unknown dilaton tadpole boundary term \cite{Erler4} to reproduce the vacuum sphere amplitude. In the Hamiltonian formalism we hope to sidestep these questions or view them in a different light, clearing the way to computation of gravitational phase space observables in string theory.

\subsection*{Acknowledgments}

TE would like to thank Daniel Grumiller, Carlo Maccaferri, Ashoke Sen, and Jakub Vo{\v s}mera for discussion of boundary terms and surface charges. AHF thanks Mukund Rangamani for conversations. We thank Igor Khavkine for comments on the manuscript and useful references to mathematics literature. TE and AHF thank David Gross for hospitality at the KITP while carrying out part of this work. The work of VB and TE was supported by the European Structural and Investment Funds and the Czech Ministry of Education, Youth and Sports (project No. FORTE—CZ.02.01.01/00/22\_008/0004632). The work of AHF is supported by the U.S. Department of Energy, Office of Science, Office of High Energy Physics of U.S. Department of Energy under grant Contract Number DE-SC0009999, and the funds from the University of California. This research was supported in part by grant NSF PHY-2309135 and the Gordon and Betty Moore Foundation Grant No. 2919.02 to the Kavli Institute for Theoretical Physics (KITP).

\begin{appendix}

\section{Tensor coalgebra and the symmetric coalgebra}
\label{app:tensor}

In this appendix we describe an isomorphism between the symmetric subspace of the tensor coalgebra $T\H$ and the symmetric coalgebra $S\H$. Both of these spaces are symmetric, but their coalgebra structures are different. This difference is reflected in certain factorial normalizations appearing in the products and homotopy algebra relations. In this paper we describe $L_\infty$ algebras using the tensor coalgebra. Though the normalizations are not standard, this has the advantage that the BV inner product can act on the tensor coalgebra without being zero. This makes it possible to discuss cyclicity in the coalgebra formalism, which is a critical part of understanding how the symplectic structure and conserved charges work. The content of this appendix was originally explained in a different way in \cite{Lada2}.

We start by defining the tensor coalgebra $T\H$. This is a direct sum of all tensor products of the vector space $\H$:
\begin{equation}
T\H = \H^{\otimes 0}\oplus \H \oplus \H^{\otimes 2}\oplus\H^{\otimes 3} \oplus \cdots,
\end{equation}
where $\H^{\otimes n}$ consists of linear combinations of states of the form
\begin{equation}A_1\otimes A_2\otimes \cdots\otimes A_n,\ \ \ \ \ A_i\in \H \, ,\end{equation}
and $\H^{\otimes 0}$ consists of scalar multiples of the identity element of the tensor product $1_{T\H}$ which satisfies
\begin{equation}1_{T\H}\otimes A = A\otimes 1_{T\H} = A,\ \ \ \ A\in\H. \end{equation}
The tensor coalgebra has a coproduct
\begin{equation}\triangle: T\H\to T\H\otimes'T\H \, , \end{equation}
which maps one copy of $T\H$ to two copies $T\H\otimes'T\H$. We use $\otimes'$ to denote the tensor product of copies of $T\H$, which is different from the tensor product $\otimes$ which defines $T\H$ itself. The specific form of the coproduct that is relevant is the {\it deconcatenation coproduct}, which acts on tensor products of states as
\begin{equation}
\triangle( A_1\otimes \cdots\otimes A_N ) = \sum_{k=0}^N A_1\otimes \cdots \otimes A_k \otimes' A_{k+1}\otimes \cdots \otimes A_N,
\end{equation}
where $k=0$ ($k=n$) the leftmost (rightmost) factor is $1_{T\H}$.  Note that $1_{T\H}$ is not the identity with respect to $\otimes'$. The coproduct is co-associative, 
\begin{equation}(\triangle\otimes' \mathbb{I}_{T\H})\triangle =(\mathbb{I}_{T\H}\otimes' \triangle)\triangle.\end{equation}
A coderivation is an operator on $T\H$ which satisfies the co-Leibniz rule
\begin{equation}\triangle {\bf D}  = ({\bf D}\otimes' \mathbb{I}_{T\H} + \mathbb{I}_{T\H} \otimes ' {\bf D})\triangle \, ,\end{equation}
where $\mathbb{I}_{T\H}$ is the identity operator on the tensor coalgebra. A group-like element satisfies 
\begin{equation}\triangle \frac{1}{1-\Phi} = \frac{1}{1-\Phi}\otimes' \frac{1}{1-\Phi} \, ,\end{equation}
which can be seen as dual to the relation defining an algebra character. From these definitions it is possible to show that coderivations and group-like elements take the form
\begin{subequations}
\begin{align}
{\bf D}( A_1\otimes \cdots\otimes A_N) & = \sum_{n=0}^N \sum_{k=0}^{N-n} (-1)^{|D_n|(|A_1|+\cdots+|A_k|)}A_1\otimes\cdots \otimes A_k \nonumber \\
& \ \ \ \ \ \ \ \ \ \ \ \ \ \ \ \ \ \ \ \ \ \ \ \ \ \ \ \ \  
\otimes D_n(A_{k+1},\!\cdots\!,A_{k+n})\otimes A_{k+n+1}\otimes\cdots\otimes A_N,\\[.5cm]
\frac{1}{1-\Phi} & = 1_{T\H} + \Phi + \Phi\otimes \Phi + \Phi\otimes \Phi \otimes\Phi + \cdots \, ,
\end{align}
\end{subequations}
where $\Phi\in\H$ is commuting and $D_n:\H^{\otimes n}\to\H$ are $n$-fold products that form the coderivation.

The symmetric coalgebra $S\H$ is the direct sum of symmetric tensor products of the vector space $\H$:
\begin{equation}
S\H = \H^{\wedge 0}\oplus \H \oplus \H^{\wedge 2} \oplus \H^{\wedge 3}\oplus \cdots,
\end{equation}
where $\H^{\wedge n}$ consists of linear combinations of elements of the form
\begin{equation}A_1\wedge A_2\wedge \cdots \wedge A_n,\ \ \ \ A_i\in \H,\end{equation}
where $\wedge$ is like the tensor product but is graded symmetric, so the order of the states does not matter.  $\H^{\wedge 0}$ consists of scalar multiples of the identity element of the symmetric tensor product
\begin{equation}
1_{S\H}\wedge A = A,\ \ \ \ A\in\H.
\end{equation}
The symmetric coalgebra has a coproduct 
\begin{equation}\triangle: S\H\to S\H\otimes'S\H \, , \end{equation}
which maps one copy of $S\H$ to two copies $S\H\otimes'S\H$. This is called the {\it unshuffle coproduct}, and takes the form 
\begin{equation}
\triangle (A_1\wedge\cdots \wedge A_N ) = \sum_{k=0}^N \sum_\sigma\frac{(-1)^\sigma}{k!(N-k)!}A_{\sigma(1)}\wedge\cdots\wedge A_{\sigma(k)}\otimes' A_{\sigma(k+1)}\wedge\cdots\wedge A_{\sigma(N)}.
\end{equation}
When $k=0$ ($k=N$) the leftmost (rightmost) factor is $1_{S\H}$. The sum is performed over all permutations $\sigma$ of the labels $1,...,N$ and $(-1)^\sigma$ is the Koszul sign derived through ordering the states as prescribed by the permutation. The factorials compensate for overcounting from permutations that only rearrange labels before and after $\otimes'$. So the nontrivial sum is over permutations that exchange labels across $\otimes'$, referred to as {\it unshuffles}. 
The coproduct is co-associative, 
\begin{equation}(\triangle\otimes' \mathbb{I}_{S\H})\triangle =(\mathbb{I}_{S\H}\otimes' \triangle)\triangle \, ,\end{equation}
where $\mathbb{I}_{S\H}$ is the identity operator on the symmetric coalgebra. A coderivation is an operator on $S\H$ which satisfies the co-Leibniz rule
\begin{equation}\triangle {\bf D}  = ({\bf D}\otimes' \mathbb{I}_{S\H} + \mathbb{I}_{S\H} \otimes ' {\bf D})\triangle \, ,\end{equation}
where $\mathbb{I}_{S\H}$ is the identity operator on the symmetric coalgebra. A group-like element satisfies 
\begin{equation}\triangle e^\Phi = e^\Phi \otimes' e^\Phi.\end{equation}
From these definitions it is possible to show that coderivations and group-like elements take the form
\begin{subequations}
\begin{align}
{\bf D} (A_1\wedge \cdots\wedge A_N) & = \sum_{n=0}^N \sum_\sigma \frac{(-1)^\sigma }{n!(N-n)!}D_n(A_{\sigma(1)},\cdots,A_{\sigma(n)})\wedge A_{\sigma(n+1)}\wedge\cdots\wedge A_{\sigma(N)},\\[.5cm]
e^\Phi & = 1_{S\H} + \Phi + \frac{1}{2!}\Phi\wedge\Phi + \frac{1}{3!}\Phi\wedge\Phi\wedge\Phi + \cdots \, ,
\end{align}
\end{subequations}
where $\Phi\in\H$ is commuting and $D_n:\H^{\wedge n}\to\H$ are $n$-fold products that form the coderivation.

The tensor coalgebra has a symmetric subspace, consisting of elements that are preserved by reorderings of tensor products. We introduce a projector onto this subspace that computes the average of all possible reorderings of the tensor product
\begin{equation}
\mathrm{Sym}( A_1\otimes\cdots\otimes A_n )= \frac{1}{n!}\sum_\sigma (-1)^\sigma A_{\sigma(1)}\otimes\cdots\otimes A_{\sigma(n)}.
\end{equation}
The coproduct, coderivations, and group-like elements are defined consistently within the symmetric subspace because
\begin{subequations}
\begin{align}
\triangle \, \mathrm{Sym} & = (\mathrm{Sym}\otimes' \mathrm{Sym})\triangle \, \mathrm{Sym},\\
{\bf D}\,\mathrm{Sym} & = \mathrm{Sym}\, {\bf D}\,\mathrm{Sym},\label{eq:DSHDTH}\\
\frac{1}{1-\Phi}& = \mathrm{Sym}\frac{1}{1-\Phi}. 
\end{align}
\end{subequations}
This means that the symmetric subspace of $T\H$ is a coalgebra by itself. The relation to the symmetric coalgebra $S\H$ is given by a kind of Borel transform 
\begin{equation}\mathcal{B}: \mathrm{Sym}(T\H)\to S\H,\end{equation}
which acts as
\begin{equation}
\mathcal{B}\,\mathrm{Sym}( A_1\otimes\cdots\otimes A_n ) = \frac{1}{n!}A_1\wedge\cdots\wedge A_n.
\end{equation}
It is possible to show that this maps the coproducts, coderivations, and group-like elements:
\begin{subequations}
\begin{align}
(\triangle_{S\H}\, \mathcal{B})\, \mathrm{Sym} & = (\mathcal{B}\otimes'\mathcal{B}\, \triangle_{T\H})\, \mathrm{Sym},\\
({\bf D}_{S\H}\, \mathcal{B})\, \mathrm{Sym} & =  (\mathcal{B}\,{\bf D}_{T\H})\, \mathrm{Sym},\\
\mathcal{B}\frac{1}{1-\Phi} & = e^\Phi.
\end{align}
\end{subequations}
The subscripts indicate whether the coproduct and coderivations as defined on $T\H$ and $S\H$. 

The products $L_n$ of a cyclic $L_\infty$ algebra define a nilpotent coderivation ${\bf L}$ on the symmetric  coalgebra,
\begin{equation}{\bf L}^2 = 0. \label{eq:Lsq}\end{equation}
Using \eq{DSHDTH},
\begin{equation}{\bf L}\, \mathcal{B} \,\mathrm{Sym}= \mathcal{B}\,{\bf M}\,\mathrm{Sym}.\label{eq:LM}\end{equation}
This can be mapped to a nilpotent coderivation ${\bf M}$ on the symmetric part of the tensor coalgebra
\begin{equation}{\bf M}^2\,\mathrm{Sym} = 0 \, . \end{equation}
An $A_\infty$ algebra is defined by an ${\bf M}$ which is nilpotent even without the symmetric projection.  By projecting onto the 1-state output of \eq{LM}, we learn that the products $L_n$ forming the coderivation ${\bf L}$ are related to the products $M_n$ forming the coderivation ${\bf M}$ as 
\begin{equation}M_n = \frac{1}{n!}L_n.\end{equation}
This explains the origin of the factorial normalizations in \eq{SHTHnorm}. When formulated in terms of the products $M_n$, it appears as though the $L_\infty$ algebra is associative up to homotopy. This is surprising because usually $L_\infty$ algebras are understood as deformations of Lie algebras, not associative algebras. The map $\mathcal{B}$ must convert the associator into a Jacobiator. Part of what makes this possible is that inputs of the associator are symmetrized. Let us explain this in more detail. To implement the symmetrization, consider homotopy associativity for a single commuting state $A$:
\begin{align}
0 =&\, QM_3(A,A,A)+M_3(QA,A,A)+M_3(A,QA,A)+M_3(A,A,QA)\nonumber\\
&\ \ \ \ \ \ \ \ \ \ \  \ \ \ \ \ \ \ \  \ \ \ \ \ \ \ \  \ \ \ \ \ \ \ \  \ \ \ \ \ \ \ \ \quad\quad +M_2(M_2(A,A),A)+M_2(A,M_2(A,A)).
\end{align} 
Since the products $M_n$ are graded symmetric, some terms above are equal and we can simplify 
\begin{equation}
0= QM_3(A,A,A)+3 M_3(QA,A,A) + 2 M_2(M_2(A,A),A).
\end{equation}
Next we express this in terms of the properly normalized $L_\infty$ products
\begin{equation}
0=\frac{1}{3!}\Big( QL_3(A,A,A)+3 L_3(QA,A,A)\Big)+ \frac{2}{(2!)^2} L_2(L_2(A,A),A).
\end{equation}
Canceling the 2s and multiplying through by $3!$ gives
\begin{equation}
0= QL_3(A,A,A)+3 L_3(QA,A,A) + 3 L_2(L_2(A,A),A).
\end{equation}
This is the homotopy Jacobi identity evaluated with a single state $A$. The $3$ multiplying the last term comes from the three terms of the Jacobiator, which happen to be equal in this case. The $3$ in the second term comes from the three ways $Q$ can act on three states, which again happen to be equal. 

\section{Conserved charge as an on-shell observable}
\label{app:cyclicity}
 
Here we elaborate on the coalgebra description of the conserved charge \eq{Fvcoalg}
\begin{equation}
F_\xi = -\int_0^1 ds\,\omega\!\left(\frac{\d\Phi(s)}{\d s}, \pi_1\Big(\M\sigma\xxi-\xxi\sigma\M\Big)\frac{1}{1-\Phi(s)}\right),
\end{equation}
and show that it is an on-shell observable. The characteristic state is 
\begin{equation}
f_\Phi = \pi_1 \Big(\M\sigma\xxi-\xxi\sigma\M\Big)\frac{1}{1-\Phi}.\label{eq:fphi2}
\end{equation}
In \eq{Qfcons} we checked that this is annihilated by $Q_\Phi$ on-shell. What remains to be shown is that the  characteristic operator $F_\Phi$ is cyclic. Since the interpolation $\Phi(s)$ can be taken off-shell, this needs to hold without assuming the equations of motion. The operator $F_\Phi$ will be cyclic if it takes the form 
\begin{equation}F_\Phi A = \pi_1\f\frac{1}{1-\Phi}\otimes A\otimes \frac{1}{1-\Phi},\end{equation}
where $\f$ is a cyclic coderivation. Unfortunately the operator $\M\sigma\xxi-\xxi\sigma\M$ is not a coderivation. But there is an equivalent operator which is:
\begin{equation}\f = [\M,\sigma \xxi]+[\sigma \M,\xxi] -\sigma[\M,\xxi].\label{eq:codf}\end{equation}
The first two terms are commutators of coderivations, and therefore are coderivations by themselves. The last term is defined as the coderivation derived from applying the sigmoid to the products contained in $[\M,\xxi]$. Note that $[\M,\xxi]$ can only be assumed to vanish acting on symmetrized elements of $T\H$, so this term in general may be nontrivial. Projecting onto the 1-state component of the tensor algebra and expanding the commutators there is some cancellation which leaves
\begin{equation}
\pi_1\f  = \pi_1\Big(\M\sigma\xxi-\xxi\sigma\M\Big).\label{eq:pi1f}
\end{equation}
Therefore $\f$ correctly defines the characteristic state of the conserved charge \eq{fphi2}. What remains to show is that $\f$ is cyclic, which can be expressed as 
\begin{equation}\langle\omega|\pi_2\f = 0.\label{eq:pi2f}\end{equation}
Substituting \eq{codf}, expanding the commutators, and noting that $\M$ and $\xxi$ annihilate $\langle\omega|$ by cyclicity, we are lead to 
\begin{equation}
\langle\omega|\pi_2\f = \langle\omega|\pi_2 \Big((\sigma \M)\xxi -(\sigma\xxi)\M -\sigma[\M,\xxi]\Big).
\end{equation}
To proceed we use the ``triangle formalism'' of \cite{Erler6}. We decompose the 2-state projector using the product and coproduct
\begin{equation}
\pi_2 = \inverttriangle (\pi_1\otimes' \pi_1)\triangle.
\end{equation}
Pulling the coproduct through the coderivations gives the expression
\begin{align} 
\langle\omega|\pi_2\f = \langle\omega|\inverttriangle (\pi_1\otimes' \pi_1) \bigg[& \Big((\sigma \M)\otimes' \mathbb{I}_{T\H} + \mathbb{I}_{T\H}\otimes\ (\sigma\M)\Big)\Big(\xxi\otimes' \mathbb{I}_{T\H} + \mathbb{I}_{T\H} \otimes' \xxi\Big)\nonumber\\ &  -\Big((\sigma\xxi)\otimes' \mathbb{I}_{T\H} + \mathbb{I}_{T\H} \otimes' (\sigma\xxi)\Big)\Big(\M \otimes' \mathbb{I}_{T\H} + \mathbb{I}_{T\H}\otimes' \M\Big) \nonumber\\
&-(\sigma[\M,\xxi])\otimes' \mathbb{I}_{T\H} + \mathbb{I}_{T\H}\otimes'(\sigma[\M,\xxi]) \bigg]\triangle .
\end{align}
Multiply out terms
\begin{align}
\langle\omega|\pi_2\f = \langle\omega|\inverttriangle \Big[& \big(\pi_1\sigma \M \xxi \big)\otimes' \pi_1 + \big(\pi_1 \sigma \M\big)\otimes' \big(\pi_1\xxi\big) + \big(\pi_1\xxi\big)\otimes'\big(\pi_1 \sigma\M\big) + \pi_1\otimes' \big(\pi_1\sigma\M\xxi\big)\nonumber\\ 
& -\big(\pi_1\sigma \xxi \M \big)\otimes' \pi_1 - \big(\pi_1\sigma\xxi\big)\otimes'\big(\pi_1 \M\big) - \big(\pi_1 \M\big)\otimes' \big(\pi_1\sigma\xxi\big) - \pi_1\otimes' \big(\pi_1\sigma\xxi \M\big)\nonumber\\
&-\big(\pi_1\sigma[\M,\xxi]\big)\otimes' \pi_1 -\pi_1\otimes'\big(\pi_1\sigma[\M,\xxi]\big)\Big]\triangle .
\end{align}
The commutator terms cancel out. There are still mixed terms where $\M$ and $\xxi$ act on different vector spaces. In these terms we factor out the sigmoid, which leads to the expression 
\begin{align}
\langle\omega|\pi_2\f = \langle\omega|\big(\sigma\otimes \mathbb{I}-\mathbb{I}\otimes \sigma)\inverttriangle\Big[\big(\pi_1\M\big)\otimes' \big(\pi_1\xxi)-\big(\pi_1\xxi\big)\otimes' \big(\pi_1\M)\Big]\triangle.
\end{align}
This vanishes because the sigmoid is preserved through the BV inner product, \eq{sigma_omega}. This shows that $\f$ is a cyclic coderivation, and therefore the characteristic operator $F_\Phi$ is also cyclic. This holds even without assuming the equations of motion. Therefore the conserved charge is an on-shell observable. However it is not an off-shell observable. We can check that 
\begin{equation}[\M,\f] = [[\M,\sigma\M],\xxi],\end{equation}
which is unfortunately not zero. We do not know if there is an off-shell modification of $\f$ that will make this commutator zero. Therefore we cannot use the conserved charge to make a gauge invariant deformation of the action.

\section{DeWitt notation}
\label{app:fermions}

In this appendix we explain how to translate our results into DeWitt notation. We introduce a basis $e_i$ for the grade zero subspace of $\H$ and a dual basis $e^i$ for the grade 1 subspace of $\H$ so that
\begin{align}
A \in \H, \ \text{grade 0}&\ \ \rightarrow \ \ A = e_i A^i,\nonumber\\
B \in \H, \ \text{grade 1}&\ \ \rightarrow\ \  B = e^i B_i.
\end{align}
where $A^i$ and $B_i$ are coefficient fields. The bases are conjugate to each other through the BV inner product
\begin{equation}\omega(e_i,e^j) = \delta_i^j.\end{equation}
Unlike \cite{Bernardes} we do not assume that all $e_i$s commute. We write the symbol $|i|$ which is even or odd if $e_i$ is commuting or anticommuting. The parity of $e^i$ must be opposite that of $e_i$. Since the dynamical field $\Phi$ is assumed to commute and can be written
\begin{equation}\Phi = e_i \phi^i,\end{equation}
the components $\phi^i$ must anticommute whenever $e_i$ anticommutes. In this way we can incorporate fermions into the formalism. 

In section \ref{sec:formalism1} and \cite{Bernardes5} we introduced a number of objects that can be expanded into components according to the table 
\begin{center}
\renewcommand{\arraystretch}{1.5}
\begin{tabular}{|cc|cc|}
\hline
$\begin{matrix}\text{Euler-Lagrange} \\[-8pt] \text{state}\end{matrix}$ & \ \ $q_\Phi = e^i q_i $\ \ \ \ & $\begin{matrix}\text{kinetic} \\[-8pt] \text{operator}\end{matrix}$  &\ \  $Q_\Phi e_i = e^j Q_{ji}$\\
\hline
$\begin{matrix}\text{generating} \\[-8pt] \text{parameter}\end{matrix}$ & $\xi_\Phi = e_i \xi^i $ & $\begin{matrix}\text{generating} \\[-8pt] \text{operator}\end{matrix}$  & $\Xi_\Phi e_i = e_j \Xi^j_i $\\
\hline
$\begin{matrix}\text{characteristic} \\[-8pt] \text{state}\end{matrix}$ & $f_\Phi = e^i f_i$ & $\begin{matrix}\text{characteristic} \\[-8pt] \text{operator}\end{matrix}$  & $F_\Phi e_i = e^j F_{ji}$\\
\hline
sigmoid & $\sigma e_i = e_j \sigma^j_i $  &  propagator & $\Delta_\Phi e^i = e_j\Delta^{ji}$ \\
\hline
\end{tabular}
\end{center}
where $\Delta$ may refer to the advanced, retarded, or causal propagators whose components we write respectively as $\Delta_A^{ij},\Delta_R^{ij},\Delta_\text{causal}^{ij}$. The causal propagator and Poisson bracket are discussed in \cite{Bernardes5}, but we include it here for completeness. The components have commuting/anticommuting parity
\begin{align}
& |q_i| = |\xi^i| = |f_i| = |i|,\nonumber\\
& |Q_{ij}| = |\Xi_i^j| = |F_{ij}| = |\sigma_i^j| = |\Delta^{ij}| = |i|+|j|.
\end{align}
Conjugation under the BV inner product implies symmetry relations 
\begin{subequations}
\begin{align}
Q_{ij}& =-(-1)^{(|i|+1)(|j|+1)}Q_{ji},\\
F_{ij}& =-(-1)^{(|i|+1)(|j|+1)}F_{ji}, \\
\Delta_R^{ij} & = (-1)^{|i||j|}\Delta_A^{ji},\label{eq:DeltaSym}
\end{align}
\end{subequations}
and defines the action of the generating operator and the sigmoid at grade 1 
\begin{equation}
\Xi_\Phi e^i = - (-1)^{|j|(|i|+1)}e^j \Xi_j^i,\ \ \ \ \sigma e^i = (-1)^{|j|(|i|+1)} e^j \sigma^i_j.
\end{equation}
The components of all objects we need can be obtained by the following recipe. We expand the action in terms of a fluctuation $\varphi^i$ of a field $\phi^i$
\begin{equation}
S[\phi^i+\varphi^i] = S[\phi^i] -\varphi^i q_i -\frac{1}{2}\varphi^i Q_{ij}\varphi^j +\cdots.
\end{equation}
This allows us to read off the components of the Euler-Lagrange state and the kinetic operator. If the symmetry transformation is known the generating parameter is determined directly by
\begin{equation}\L_\xi \phi^i = \xi^i,\end{equation}
while the generating operator is found by calculating the exterior derivative,
\begin{equation}\deltahat \xi^i = (-1)^{|i|+|j|}\Xi^i_j \delta\phi^j.\end{equation}
The characteristic state and the characteristic operator of an observable can be determined in the same way as the Euler-Lagrange state and kinetic operator from the action, by expanding in terms of a fluctuation $\varphi^i$ of a field $\phi^i$.
The retarded propagator can be found by solving the equation
\begin{equation}Q_{ij}\Delta_R^{jk}f_k = f_i,\label{eq:QDeltaf}\end{equation}
where $f_i$ are the components of the characteristic state of an observable, subject to the condition that $\Delta_R^{ij}f_j$ vanishes (fast enough) in the infinite past. The advanced propagator is similarly determined with the condition that $\Delta_A^{ij}f_j$ vanishes in the infinite future. The causal propagator is 
\begin{equation}
\Delta^{ij}_\text{causal} = \Delta^{ij}_R-\Delta^{ij}_A.
\end{equation}
With these definitions it is possible to translate all formulas to DeWitt notation:
\begin{subequations}
\begin{align}
& \text{Action:}\ \ \ \ \ \ \ \ \ \ \ \ \ \ \ \ \ \ \ \ \ \ \ \ \ \ \ \ \ S  = -\int_0^1 ds\, \frac{\d \phi^i(s)}{\d s} q_i(s),\\
& \text{Observable:}\ \ \ \ \ \ \ \ \ \ \ \ \ \ \ \ \ \ \ \ \ \ \ F  = -\int_0^1 ds\, \frac{\d \phi^i(s)}{\d s} f_i(s),\\
& \text{Symplectic form:}\ \ \ \ \ \ \ \ \ \ \ \ \ \ \ \ \Omega  = \frac{1}{2}(-1)^{|j|+1}\delta\phi^i\Big[Q_{ik}\sigma^k_j - (-1)^{|i|(|k|+1)}\sigma_i^k Q_{kj}\Big] \delta\phi^j,\\
&\!\begin{matrix}\text{Conserved charge }\\ \text{(localized form):\ \ \  }\end{matrix}\ \ \ \ \ \ \ \ \ \ \ \ \  F_\xi = -\int_0^1 ds \frac{\d\phi^i(s)}{\d s}\Big[(-1)^{|i|(|j|+1)+|j|(|k|+1)}\big(\Xi_i^j(s)\sigma_j^k-\sigma_i^j\Xi_j^k(s)\big)q_k(s)\nonumber\\
&  \ \ \ \ \ \ \ \ \ \ \ \ \ \ \ \ \ \ \ \ \ \ \ \ \ \ \ \ \ \ \ \ \ \ \ \ \ \ \ \ \ \ \ \ \ \ \ \ \ \ \ \ \ \ \ \ 
+\big(Q_{ij}(s)\sigma^j_k -(-1)^{|i|(|j|+1)}\sigma_i^jQ_{jk}(s)\big)\xi^k(s)\Big],\label{eq:Fv_comp}\\
&\!\begin{matrix}\text{Conserved charge\ \ \ \ \ \ \ }\\ \text{(total derivative form):}\end{matrix}\ \ \ \ \ \ \ F_\xi = - (-1)^{|i|(|j|+1)}\xi^i\sigma^j_i q_j\big|_\tau\label{eq:Fvtau_comp},\\[5pt]
& \text{Poisson bracket:}\ \ \ \ \ \ \ \ \ \ \ \ [F,G] = (-1)^{|i|+1}f_i\Delta^{ij}_{\text{causal}}g_j,
\end{align}
\end{subequations}
where $q_i(s)$ and so on are components evaluated on the interpolating field $\phi^i(s)$ and the symbol $|_\tau$ in \eq{Fvtau_comp} indicates that all instances of $\phi^i$ come with tau regularization. 

\end{appendix}

\end{document}